\documentclass[aps,prl,twocolumn,superscriptaddress,longbibliography,nobalancelastpage]{revtex4-2}%
\usepackage{graphicx}
\usepackage{float}
\usepackage{dcolumn}
\usepackage{dutchcal}
\usepackage{bm,color}
\usepackage{amsmath,amssymb,dsfont,amstext,amsfonts}
\usepackage{extarrows}
\usepackage[colorlinks=true,linkcolor=blue,urlcolor=blue,citecolor=blue]{hyperref}
\usepackage{xcolor}
\usepackage{wasysym}
\usepackage{mathtools}
\usepackage{bbold} 
\usepackage{braket}
\usepackage{dsfont}
\usepackage{bigints}
\usepackage{soul}

\usepackage{comment}
\usepackage{environ}

\newif\ifSM

\SMtrue

\ifSM
\NewEnviron{supplementalMaterials}{%

	\setcounter{figure}{0}
	\setcounter{equation}{0}
	\setcounter{page}{1}
	\BODY
}
\else
\excludecomment{supplementalMaterials}
\fi

\makeatletter
\def\maketitle{
	\@author@finish
	\title@column\titleblock@produce
	\suppressfloats[t]}
\makeatother

\hypersetup{
	colorlinks,
	linkcolor={red!95!black},
	citecolor={green!60!black},
	urlcolor={blue!95!black}
}

\definecolor{Nathanblue}{rgb}{0.,0.24,0.51}

\usepackage[normalem]{ulem} 
\usepackage{color}

\newcommand{\approptoinn}[2]{\mathrel{\vcenter{
  \offinterlineskip\halign{\hfil$##$\cr
    #1\propto\cr\noalign{\kern2pt}#1\sim\cr\noalign{\kern-2pt}}}}}

\newcommand{\appropto}{\mathpalette\approptoinn\relax}

\newcommand{\nodagger}{{\vphantom{\dagger}}}
\newcommand{\cll}{\text{$\chi$LL} }

\newcommand{\br}{\bm{r}}

\def\Xint#1{\mathchoice
	{\XXint\displaystyle\textstyle{#1}}%
	{\XXint\textstyle\scriptstyle{#1}}%
	{\XXint\scriptstyle\scriptscriptstyle{#1}}%
	{\XXint\scriptscriptstyle\scriptscriptstyle{#1}}%
	\!\int}
\def\XXint#1#2#3{{\setbox0=\hbox{$#1{#2#3}{\int}$}
		\vcenter{\hbox{$#2#3$}}\kern-.5\wd0}}

\def\dashint{\Xint-}

\makeatletter
\renewcommand{\@fnsymbol}[1]{}
\makeatother

\begin{document}

\title{\color{Nathanblue}Circular dichroism on the edge of quantum Hall systems:\\~From many-body Chern number to anisotropy measurements}



\author{F.~Nur \"{U}nal}
\email{fnu20@cam.ac.uk}
\affiliation{School of Physics and Astronomy, University of Birmingham, Edgbaston, Birmingham B15 2TT, United Kingdom\looseness=-1}
\affiliation{TCM Group, Cavendish Laboratory, University of Cambridge, JJ Thomson Avenue, Cambridge CB3 0HE, United Kingdom\looseness=-1}

\author{A. Nardin}
\email{alberto.nardin@universite-paris-saclay.fr}
\affiliation{Universit\'e Paris-Saclay, CNRS, LPTMS, 91405 Orsay, France}

\author{N. Goldman}
\email{nathan.goldman@lkb.ens.fr}
\affiliation{Laboratoire Kastler Brossel, Coll\`ege de France, CNRS, ENS-Universit\'e PSL, Sorbonne Universit\'e, 11 Place Marcelin Berthelot, 75005 Paris, France}
\affiliation{International Solvay Institutes, 1050 Brussels, Belgium}
\affiliation{Center for Nonlinear Phenomena and Complex Systems, Universit\'e Libre de Bruxelles, CP 231, Campus Plaine, B-1050 Brussels, Belgium}

\date{\today}

\begin{abstract}

{ Quantum Hall states are characterized by a topological invariant, the many-body Chern number, which determines their quantized Hall conductivity. This invariant also emerges in circular dichroic responses, namely, by applying a circular drive and comparing excitation rates for opposite orientations. This work explores the dichroic response of confined, isolated quantum Hall systems, where bulk and edge contributions cancel exactly:~When the edge response is properly isolated, the circular dichroic signal becomes quantized, serving as a direct and elegant probe of the many-body Chern number encoded in the edge physics. We demonstrate that this quantized edge response is entirely captured by low-energy chiral edge modes, allowing for a universal description of this effect based on Wen's edge theory. Its low-energy nature implies that the quantized edge response can be distinguished from the bulk response in the frequency domain. The edge response is also shown to be a sensitive diagnostic of geometric features. This opens the possibility of characterizing the shape of quantum Hall droplets through edge spectroscopic measurements, without requiring knowledge of the system's  boundary profile. We illustrate our findings using realistic models of integer and fractional Chern insulators, with different edge geometries, and propose detection schemes suitable for ultracold atoms. }

\end{abstract}

\maketitle
  
{\it Introduction---} Since the discovery of the quantum Hall (QH) effects~\cite{girvin2002quantum}, the robust quantization of macroscopic  observables has been recognized as the manifestation of topology in physics~\cite{Thouless_KNN,Thouless_invariant,volovik2003universe,bernevig2013topological}, sparking tremendous attention in both solid state~\cite{Hasan_RevModPhys,Qi_RevModPhys} and quantum engineered settings~\cite{Cooper19_RMP,ozawa2019topological}.
This vast exploration of topological states has revealed various quantized effects, accessible through a rich variety of experimental probes, such as transport, interferometry, spectroscopy and tomography~\cite{Hasan_RevModPhys,Qi_RevModPhys,Cooper19_RMP,ozawa2019topological}.

Coupling light to topological matter can lead to novel phenomena~\cite{schlawin2022cavity,bloch2022strongly,bacciconi2024theory}, and in particular, to quantized optical responses~\cite{wu2016quantized,de2017quantized,Tran17_SciAdv_dichrosim,ahn2022riemannian,kruchkov2023spectral,souzaVanderbilt08_PRB,Morimoto09_PRL, Fregosa18_PRL_jerk, bouhon2023plucker,onishi2024fundamental,jankowski2023optical}. As a prominent example, circular dichroism (CD) has been proposed as a powerful probe for extracting many-body topological invariants and quantum geometry in a broad class of topological systems~\cite{Tran17_SciAdv_dichrosim,OzawaGoldman18_PRB_metric,Repellin_dichroism,Klein_dichroism,Hur_dichroism,Pozo_dichroism,Ozawa_MB_quantum_geometry,Werner_dichroism,schuler2020local,dichroism_Josephson,Molignini_dichroism,goldman2024relating,jankowski2023optical,kim2023circular}. This approach has been demonstrated in ultracold atoms, where the CD response was measured to evaluate the Chern number of Floquet-engineered Bloch bands~\cite{Asteria19_NatPhys}; it was also applied in solid-state qubit systems to extract the full quantum geometric tensor~\cite{yu2020experimental,QGT_SC,yu2022quantum,yu2024experimental} and to reveal exotic monopoles~\cite{chen2022synthetic,tensor_monopole_SC}. The quantized CD response of strongly-correlated fractional Chern insulators (FCIs), which reflects the underlying many-body Chern number~\cite{goldman2024relating}, was also demonstrated through numerical studies~\cite{Repellin_dichroism}.

Crucially, the relation between the many-body Chern number of a 2D Chern insulator (CI) and its quantized CD response implicitly assumes that the measured response entirely emanates from the bulk~\cite{Souza_dichroic,Tran17_SciAdv_dichrosim,TranCooperGoldman18_PRA_circdichroLL,goldman2024relating}. { In fact, in a confined and isolated system (e.g.~a quantum gas in a trap), the edge contribution to the CD response exactly cancels the quantized bulk response [Fig.~\ref{fig:sketch}(a)]; this effect can be attributed to the vanishing net conductivity of closed settings. Identifying schemes that isolate the bulk response~\cite{Tran17_SciAdv_dichrosim} has thus been essential in this quantized CD framework.}

In this work, we follow a different approach and set the focus on the quantized CD response associated with the edge of quantum Hall systems. 
We demonstrate that this quantized CD response is entirely captured by low-energy edge modes, as described by Wen's chiral Luttinger liquid ($\chi$LL) theory~\cite{Wen_PRL_1990,Wen_edge_theory,Wen_AdvPhys_1995,Chang_RMP_2003}. This result has multiple consequences:~(i) the  edge CD response is universal, it depends only on the specific topological order, and can be extracted from the edge of any QH state; (ii) the bulk and edge CD responses, which are both quantized in insulating states, can be individually isolated by resolving the dichroic response in the frequency domain [Fig.~\ref{fig:sketch}(b)]; (iii) { the edge response is also a sensitive diagnostic of geometric features, which can be used to probe the shape of QH droplets}. We argue that these quantized responses can be individually measured in cold-atom experiments~\cite{Asteria19_NatPhys}, in the presence of well-designed trapping potentials~\cite{navon2021quantum,Binanti_edge}. We illustrate our findings by studying realistic models of Chern insulator  and fractional Chern insulator  states, considering different edge geometries and edge-mode configurations.


{\it Quantized CD: bulk versus edge contributions ---} We consider a general class of Chern insulators characterized by the many-body Chern number, $C_{\rm MB}$, { which can be integer or fractional depending on the nature of the state~\cite{Thouless_invariant}}. Such gapped many-body states exhibit a quantized Hall response, $\sigma_{xy}/\sigma_0\!=-C_{\rm MB}$, where $\sigma_0$ denotes the conductivity quantum~\cite{Thouless_invariant}. Interestingly, one can relate this transport coefficient to a dissipative optical response, using Kramers-Kronig relations~\cite{Stern_Faraday}:~$\sigma_{xy}=(2/\pi) \int_0^{\infty}\omega^{-1} \sigma^I_{xy}(\omega) {\rm d}\omega$. Here, the optical Hall conductivity $\sigma^I_{xy}(\omega)$ can be expressed in terms of  excitation rates upon circular driving~\cite{Tran17_SciAdv_dichrosim}
\begin{equation}
\sigma^I_{xy}(\omega)=\hbar \omega (\Gamma_+-\Gamma_-)/8 A E^2 ,
\end{equation}
where $E$ is the strength of the perturbing electric field $\mathbf{E}_\pm(\br,t)\!=\!E\left(\hat{e}_x \pm i\hat{e}_y \right)e^{-i \omega t} + \text{c.c.}$, $\Gamma_{\pm}(\omega)$ are the excitation rates associated with polarization $\pm$ and frequency $\omega$, and $A$ is the system's area. Combining these results indicates that the many-body Chern number can be directly related to integrated excitation rates~\cite{Tran17_SciAdv_dichrosim,Repellin_dichroism,goldman2024relating}
\begin{equation}   \label{eq:C_dichroism}
    \frac{\hbar^2}{2A E^2} \left( \Gamma^{\text{int}}_+ - \Gamma^{\text{int}}_-  \right) = C_{\rm MB},
\end{equation}
where we introduced the notation $\Gamma^{\text{int}}_{\pm}=\int_0^{\infty}\Gamma_{\pm}(\omega) {\rm d}\omega$. Quantized CD [Eq.~\eqref{eq:C_dichroism}] was measured in quantum engineered systems, where the excitation rates upon circular driving can be finely monitored, e.g.~using band-mapping techniques~\cite{Asteria19_NatPhys}. This quantization law was also verified numerically for FCIs~\cite{Repellin_dichroism}.


{ Crucially, Eq.~\eqref{eq:C_dichroism} pertains to a purely \emph{bulk} response and implicitly disregards any contributions to the excitation rates originating from the edge region.} In fact, for any isolated and confined system, the edge and bulk responses perfectly cancel each other~\cite{Souza_dichroic,Tran17_SciAdv_dichrosim,TranCooperGoldman18_PRA_circdichroLL}
\begin{equation}   \label{eq:C_dichroism_zero}
    \left( \Gamma^{\text{int}}_+ - \Gamma^{\text{int}}_-  \right)_{\text{edge}} = - \left( \Gamma^{\text{int}}_+ - \Gamma^{\text{int}}_-  \right)_{\text{bulk}},
\end{equation}
which can be traced back to the vanishing Hall conductivity of any isolated setting with boundaries. For a non-interacting system, this can be appreciated by noting that the CD response in Eq.~\eqref{eq:C_dichroism} can be expressed in terms of the local (real-space) Chern marker~\cite{Bianco,Tran17_SciAdv_dichrosim}
\begin{equation}   \label{eq:C_dichroism_local}
    \frac{\hbar^2}{2E^2}\left( \Gamma^{\text{int}}_+ - \Gamma^{\text{int}}_-  \right)= \, \sum_{\bf r} \, C_{{\bf r}}.
\end{equation}
As illustrated in Fig.~\ref{fig:sketch}(a), the bulk contributes to the quantized response in Eq.~\eqref{eq:C_dichroism}, $\sum_{\text{bulk}} C_{{\bf r}}\!=\!A C_{\rm MB}$, which is exactly compensated by the edge response, $\sum_{\bf r} \, C_{{\bf r}}=0$. 
{ In what follows, we rigorously establish that the edge CD response can be selectively isolated, thereby providing direct access to the many-body Chern number, an insightful manifestation of the bulk-edge correspondence.}

\begin{figure}
  \includegraphics[width=1\linewidth]{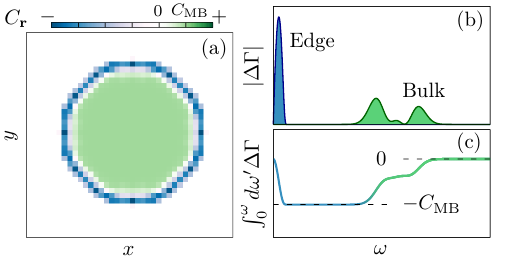}
    \caption{(a) The dichroic response of a QH state is quantized in the bulk according to the many-body Chern number $C_{\rm{MB}}$ [Eq.~\eqref{eq:C_dichroism}]. In a confined system, this bulk response is perfectly cancelled by the edge, as evidenced by the local Chern-marker description [Eq.~\eqref{eq:C_dichroism_local}]. (b) The bulk and edge  dichroic responses can be resolved in the frequency domain. (c) The response associated with low-energy edge modes is quantized, reflecting the many-body Chern number $C_{\rm{MB}}$ of the topological insulating state.}
    \label{fig:sketch}
\end{figure} 

{\it Quantized CD as a low-energy edge response ---} 
The latter { discussion} brings us to further investigate the CD response associated with the edge region, the nature of which is a priori not obvious. 
{ Although, in principle,
excitations localized at the edge can potentially occur at high energies (of the order of the bulk gap), we demonstrate that the edge CD response arises exclusively from low-energy, gapless modes localized at the system's boundary;} see Fig.~\ref{fig:sketch}(b,c).
The quantized edge dichroism can indeed be understood from the viewpoint of the \cll theory in the large particle-number ($N\gg1$) limit. For simplicity, we first consider a QH droplet with a circular boundary and a single chiral edge mode, such as an integer QH state~\cite{VonKlitzing_PRL_1980,Laughlin_PRB_1981} 
or a Laughlin-type fractional QH state~\cite{Laughlin_PRL_1983} at filling fraction $\nu\!=\!-C_{\rm MB}$ in the lowest Landau level. { Generalizations to more sophisticated settings with multiple edge modes and anisotropic boundaries will be discussed in the following paragraphs.}

{ Using Fermi's golden rule, we write the integrated rates entering Eq.~\eqref{eq:C_dichroism_zero} as
$(\Gamma_{\pm}^{{\rm int}})_{\rm edge}\!=\!2\pi E^2 S_{\pm}/\hbar^2 $, where the structure factor $S_{\pm}$ is given by~\cite{Tran17_SciAdv_dichrosim,TranCooperGoldman18_PRA_circdichroLL,stamper2002spinor,SI}}
\begin{align}
    \label{eq:edge_modes_matrix_elements}
    S_{\pm}=\sum_{n\in\text{edge}} \left| \int   (x \pm iy) \Braket{n\left| \hat{\rho}(\mathbf{r}) \right|0} d{\bf r}\right|^2  
\!= \sum_{n\in\text{edge}} M^{n,0}_{\pm} .
\end{align}
Here, $\hat{\rho}({\bf r})$ is the particle density operator, $\ket{0}$ denotes the ground state and the sum runs over the set of low-energy  { edge} modes~\cite{Wen_AdvPhys_1995}. Since the bulk is incompressible, contributions to $S_{\pm}$ can only come from a thin layer at the edge of the system, at radius $R$, with a typical thickness set by the magnetic length. We can thus approximate
$S_{\pm} \simeq R^2 \sum_n \left| \Braket{n \left| \hat{\rho}_{\pm1} \right|0} \right|^2$, where $\hat{\rho}_{\ell} = \int  e^{i\ell \theta} \hat{\rho}_{\text{eff}}(\theta) d\theta$ is the angular Fourier transform of a one-dimensional edge density, $\hat{\rho}_{\text{eff}}(\theta)=\int \hat{\rho} (\mathbf{r}) r dr$.
Following Wen's bosonization~\cite{Wen_edge_theory,Wen_AdvPhys_1995, Nardin_PRA_2023,nardin2023refermionized}, we identify $\hat{\rho}_{+\ell}=\sqrt{|C_{\rm MB}| \ell}\, \hat{b}_\ell^\dagger$,
with the operator $\hat{b}_\ell^{\dagger}$ creating a bosonic excitation along the edge of the system, and
$\hat{\rho}_{-\ell}^\nodagger=\hat{\rho}_{+\ell}^\dagger$ as its annihilation; here we explicitly used the relation between the filling fraction and the many-body Chern number. The ground state $\ket{0}$ is naturally interpreted  as the vacuum of bosonic excitations and the excited states $\ket{n}$ as the states that can be obtained from it through the action of $\hat{b}^\dagger_\ell$.
Since the edge mode carries a definite chirality, one obtains $S_{+}\!=\!|C_{\rm MB}| A/\pi$, while the opposite orientation yields $S_{-}\!=\!0$; 
here and from now on we assume $C_{\rm MB} < 0$~\cite{SI}.
As anticipated, we exactly recover Eqs.~\eqref{eq:C_dichroism}-\eqref{eq:C_dichroism_zero}, where the sum rule is entirely exhausted by the gapless edge mode. { As a consequence, edge and bulk CD responses can be distinguished via frequency selection, as we explicitly demonstrate below.}

\begin{figure}[t]
  \includegraphics[width=1\linewidth]{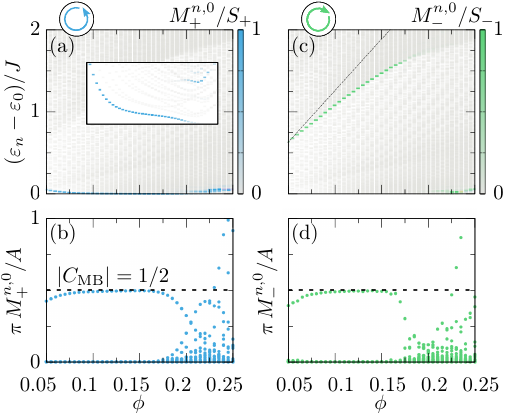}
    \caption{
    (a) Energy spectrum for $N\!=\!2$ hard-core bosons on a $12\times12$ square lattice, in the presence of a weak harmonic confinement of strength $U_h\!=\!3.5\times10^{-3} J$, as a function of the flux $\phi$.
    The points have been colored according to the magnitude of the matrix elements $M^{n,0}_+$. 
    The inset shows a close-up on the low-energy features.
    (b) The same matrix elements, without energy resolution, and compared to the theory prediction (black dashed line).
    (c,d) Same but for the matrix elements $M^{n,0}_-$, associated with the bulk response. Black dashes in (c) show the continuum-limit cyclotron gap $\hbar\omega_c/J\!=\!4\pi\phi$.
    }
    \label{fig:FCI_z}
\end{figure}

{ Having established that a quantized CD edge response is expected at low frequencies, we now illustrate this result by considering a bosonic Laughlin-type FCI state at filling factor $\nu\!=\!1/2$, in the Harper-Hofstadter model with hard-core interactions~\cite{Sorensen,Hafezi_FQH,Palmer_FQH,Moore_edge_FCI,moller2009composite,Scaffidi_Hof,Roy_Hofstadter,Moller_higher,Murthy_Hof,Raciunas_18_PRA,Repellin_PRA,Palm_PRB,Bosl}. Here, we add a weak harmonic potential $V_{m,n}\!=\!U_h (m^2+n^2)$, which helps isolating a well-defined edge-mode branch~\cite{Moore_edge_FCI,Luo_PRB_2013,Binanti_edge}, and stabilizes a Laughlin-like state with a circular edge.} 
{ We study the CD response of this FCI state by calculating the coupling matrix elements $M_{\pm}^{n,0}$ in Eq.~\eqref{eq:edge_modes_matrix_elements} for all (bulk and edge) eigenstates $\vert n \rangle$; we show our results in Fig.~\ref{fig:FCI_z} as a function of the magnetic flux $\phi$.
In agreement with the \cll prediction, we find $M_+^{n,0}\!\ne\!0$ and $M_{-}^{n,0}\!=\!0$ at low energy for a broad range of flux [Fig.~\ref{fig:FCI_z}(a,c)]. Since a single edge mode couples to the CD probe, the corresponding matrix element is directly related to the many-body Chern number $|C_{\rm MB}|\!=\!1/2$, as we demonstrate in Fig.~\ref{fig:FCI_z}(b). We note that this result can be used to identify the flux window within which the FCI state is stabilized. From Fig.~\ref{fig:FCI_z}(a), we also find that the energy of the relevant low-energy mode decreases with increasing magnetic flux, as expected from a semiclassical description of the edge.} 
{ In contrast, the probe with negative orientation couples to a single high-energy (bulk) mode, located approximately at the continuum cyclotron energy $\hbar\omega_c/J \!=\!4\pi \phi$ [Fig.~\ref{fig:FCI_z}(c)]. In accordance with Eq.~\eqref{eq:C_dichroism_zero}, the corresponding matrix element is also related to the many-body Chern number $|C_{\rm MB}|$, hence exactly compensating the low-energy edge response. We note that it is the cyclotron gap (and not the many-body gap, which is much smaller) that sets the energy scale at which absorption of opposite angular-momentum quanta occurs~\cite{SI}.}

{ This example illustrates how the bulk gap naturally induces a frequency separation, allowing isolation of the quantized CD edge response from the bulk; We performed similar calculations  
on other models~\cite{SI}, further highlighting the broad application of the result.}

\begin{figure}
  \includegraphics[width=1\linewidth]{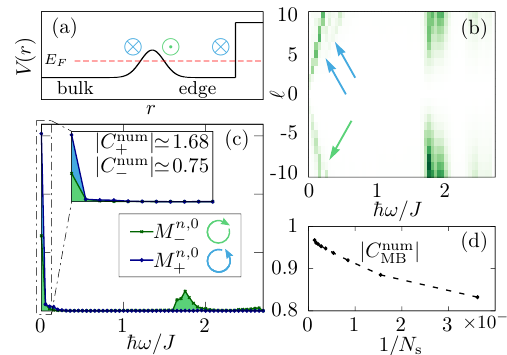}
    \caption{(a) Multiple edge modes are introduced in a non-interacting integer quantum Hall state in Harper-Hofstadter bands by including, in addition to a circular hard wall at $r_{\text{wall}}\!=\!14.5$, a ring-shaped potential bump of strength $0.5J$, centered two sites away from the walls.
    (b) Angular-momentum-resolved spectroscopy:~Matrix elements $M_{\ell}^{n,0}\!=\!\left|\int d^2\mathbf{r}\,  e^{i \ell \theta}\braket{n|\hat{\rho}(\mathbf{r})|0}\right|^2$, as a function of frequency, $\omega_{n,0}\!=\!(\varepsilon_n-\varepsilon_0)/\hbar$, and transferred (optical) angular momentum $\ell$; $\varepsilon_n$ is the energy of the $n$-th eigenstate.    
    The presence of multiple edge modes is highlighted by the arrows.
    (c) CD matrix elements $M_{\pm}^{n,0}$ [Eq.~\eqref{eq:edge_modes_matrix_elements}], as a function of $\omega_{n,0}$, with the inset zooming in on the low-energy features. The numerically extracted Chern number contributions, $C_{\text{MB}}^{\rm num} = C_+^{\rm num}-C_-^{\rm num}$, are calculated by isolating the low-energy response ($\hbar\omega\!\leq\!1J$).  The matrix elements are averaged over a small frequency window (of width $0.05J/\hbar$) for visualization. (d) $C_{\rm MB}^{\rm num}$ approaches the quantized value for increasing system size, denoted by the total number of occupied sites, $N_{\text{s}}$.
    We used $\phi\!=\!2/7$ and $\varepsilon_F\!=\!-1.5J$.
    }
    \label{fig:CI_OppositeChirality}
\end{figure}

{\it Multiple edge modes ---} 
{ The perfect edge dichroism ($S_+\!\ne\!0$ and $S_{-}\!=\!0$) discussed earlier is a distinctive result of a single edge mode circulating along a circular boundary. While the result $S_-\!=\!0$ typically fails when multiple edge channels are present,  our $\chi$LL calculations predict that the difference $S_{+}-S_{-}\!=\!|C_{\rm MB}| A/\pi$ is robustly quantized~\cite{SI}, consistent with Eqs.~\eqref{eq:C_dichroism}-\eqref{eq:C_dichroism_zero}. This multi-edge-mode scenario is relevant for hierarchy~\cite{Haldane_PRL_1983,Wen_AdvPhys_1995} or bilayer~\cite{Halperin_HelvPhysAct_1983} QH states, and more generally, in the presence of edge reconstruction~\cite{Wan_PRL_2002,Wan_PRB_2003,Yang_PRL_2003}. Here, we illustrate this key result by considering:~(i) A non-interacting CI state with reconstructed edges displaying three counter-propagating edge modes; and (ii) a strongly-interacting two-component bosonic Halperin state displaying two co-propagating edge modes~\cite{Cooper_2008,Wen_AdvPhys_1995}.}

{ We first consider a CI of non-interacting fermions on the Harper-Hofstadter model, at flux $\phi\!=\!2/7$ per plaquette, confined in a circular box. By setting the Fermi energy in the lowest gap, the CI state exhibits a single edge mode ($|C_{\rm MB}|\!=\!1$). We then add a small Gaussian potential bump, in the vicinity of the edge, in view of generating additional (counter-propagating) edge modes; see the sketch in Fig.~\ref{fig:CI_OppositeChirality}(a). We confirm the appearance of two additional edge modes by performing angular-momentum resolved spectroscopy on the edge of the QH system~\cite{Goldman_LG_2012,Binanti_edge}, as we show in Fig.~\ref{fig:CI_OppositeChirality}(b). We then represent the CD matrix elements $M_{\pm}^{n,0}$, as a function of frequency $\omega_{n,0}\!=\!(\varepsilon_n-\varepsilon_0)/\hbar$, in Fig.~\ref{fig:CI_OppositeChirality}(c). 
Given the circular shape of the boundary, the two strong signals observed at low energy, \(M_{+}^{n,0}\!\sim\!M_{-}^{n,0}\), indicate the presence of counter-propagating edge modes. We verify that their difference approaches the quantized result, $S_{+}-S_{-}\!=\!|C_{\rm MB}| A/\pi$, as one increases the system size; see Fig.~\ref{fig:CI_OppositeChirality}(d).

}

{ We then consider the incompressible bosonic Halperin $(2,2,1)$ state~\cite{Cooper_2008}, which appears in the continuum as the densest ground state for a strongly-interacting two-component system under a magnetic field~\cite{SI}, at angular momentum $L_{\rm (2,2,1)}\!=\!N_A(N_A-1)+N_B(N_B-1)+N_A N_B$ with  $N_{A/B}$ the number of bosons. 
This incompressible state has a bulk filling fraction $\nu\!=\!|C_{\rm MB}|\!=\!2/3$, and supports two co-propagating edge modes along its boundary~\cite{Wen_AdvPhys_1995}.
We numerically study a system of $N_A\!=\!5$ and $N_B\!=\!6$ particles using exact diagonalization, and consider contact two-body interactions, $V_{12}\!=\!g \delta^{(2)}(r_1-r_2)$.
Figure~\ref{fig:sm_halperin_low}(a) displays the radial density profile, which exhibits a clear plateau at \(\rho_{0}\!=\!\frac{|C_{\rm MB}|}{2\pi l_B^2}\) ($l_B\!=\!\sqrt{\hbar/qB}$ denotes the magnetic length), indicating the formation of an incompressible bulk state. Figure~\ref{fig:sm_halperin_low}(b) illustrates the many-body energy spectrum as a function of the total angular momentum, highlighting a non-degenerate ground state at \(L = L_{(2,2,1)}\) close to zero energy, and the emergence of low-energy excitations for \(L > L_{(2,2,1)}\). Note that we included an anharmonic ring-shaped box potential to make the edge modes dispersive~\cite{CooperSimon_PRL_2015, Macaluso_PRA_2017}.} 
{ A detailed examination of the matrix elements \(M_{\pm}^{n,0}\) reveals the edge-mode contributions to the dichroic response. Below the bulk energy gap, the matrix elements \(M_{-}^{n,0}\!\approx\!0\) remain negligible~\cite{SI}. 
On the other hand, as we show in Fig.~\ref{fig:sm_halperin_low}(b), $M_{+}^{n,0}\!\neq\!0$ exclusively for the two co-propagating low-energy edge modes located at $L-L_{(2,2,1)}\!=\!1$, consistent with the angular-momentum transferred by the circular probe.
One verifies that the sum of these two matrix elements  yields the many-body Chern number of the underlying Halperin state, $|C_{\rm MB}|\!=\!2/3$, hence validating our theory in this strongly-interacting multi-mode setting.}


\begin{figure}
\includegraphics[width=1\linewidth]{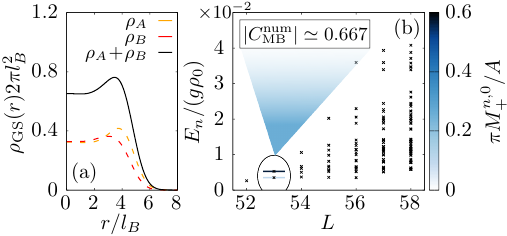}
    \caption{(a) Radial density profile $\rho_A$(/$\rho_B$) of the $A$(/$B$) component of a Halperin $(2,2,1)$ state (yellow(/red) dashes), and the combined density $\rho_A+\rho_B$ (black solid line). The bulk's density approaches $\rho_0=\nu/(2\pi l_B^2)$, with $\nu=|C_{\rm MB}|=2/3$.
    (b) Low-lying portion of the  system's spectrum $E_n$ (black crosses) as a function of its angular momentum $L$; horizontal colored lines display the $M_+^{n,0}$ matrix elements; both  edge modes at $L=L_{(2,2,1)}+1$ can be seen to carry a non-zero spectral weight, which together add-up to the many body Chern number $C_{\rm MB}$ (inset).
    }
    \label{fig:sm_halperin_low}
\end{figure}

{\it Anisotropic droplets ---} { The quantized nature of the edge CD response is not restricted to systems with circular symmetry. Indeed, a generalization of our \cll calculation demonstrates that the prediction $S_{+}-S_{-}\!=\!|C_{\rm MB}| A/\pi$ holds for arbitrary edge geometries~\cite{SI}. 

Interestingly, and quite generally, the ratio $S_{-}/S_{+}$ provides a direct measure of the anisotropy of the droplet. For instance, in the experimentally relevant case of an elliptic trap~\cite{mukherjee2022crystallization}, $V\!=\!V(e^{2\lambda}x^2+e^{-2\lambda}y^2)$, introducing an anisotropic edge, this ratio is given by~\cite{SI}
\begin{equation}
    \label{eq:SmSp_ratio_ellipse}
    S_-/S_+ = \tanh^2(\lambda).
\end{equation}
 We have verified this relation numerically, both for lattice and continuum models, for integer and fractional QH states, considering different types of confinement~\cite{SI}. Figure~\ref{fig:elliptic_laughlin} illustrates such a result for a Laughlin liquid trapped by an anisotropic (elliptic) anharmonic trap: in panel (a) we plot the ground state density; in (b) we verify the edge CD sum rule, while in (c) we demonstrate that the ratio $S_-/S_+$ gives access to the boundary's shape through Eq.~\eqref{eq:SmSp_ratio_ellipse}.

{ A natural question arises: is it possible to distinguish the ``intrinsic" contribution to $S_{\pm}$, stemming from counter-propagating modes, from the ``extrinsic" contribution induced by boundary anisotropies? We argue that, at least in principle, the latter can be mitigated by modifying the geometry of the probe, e.g.~replacing a circular probe with an elliptic one. In contrast, the intrinsic component should be robust against such geometric tuning. We leave a deeper investigation of this intriguing distinction for future work.}

Summarizing, our results demonstrate that the edge CD response is not only a robust probe of topological invariants but also a sensitive diagnostic of geometric features. In experimental settings, this opens the possibility of characterizing the shape and symmetry of QH droplets through edge spectroscopic measurements, without requiring full knowledge of the boundary profile.}

\begin{figure}
\includegraphics[width=1\linewidth]{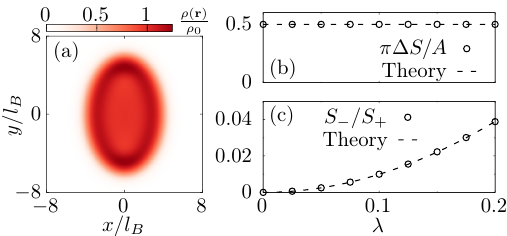}
    \caption{ 
    (a) Density $\rho(x,y)$ of a Laughlin liquid in an anisotropic (elliptic) anharmonic trap, for a fixed ellipticity parameter $\lambda=0.2$.
    (b) The two normalized rates $S_\pm$ (black circles) saturate the edge CD sum rule for all the ellipticities $\lambda$ considered, with the expected value $|C_{\rm MB}|=1/2$ (black dashes).
    (c) The ratio $S_-/S_+$ reveals $\lambda$ via Eq.~\eqref{eq:SmSp_ratio_ellipse}.
    }
\label{fig:elliptic_laughlin}
\end{figure}

{ {\it Local edge spectroscopy---}} We finally comment on the possibility of exciting the system with a generalized CD probe described by a time-dependent potential of the form $U_\pm(\mathbf{r},t)\!=\!E f(r) e^{i(\pm\theta-\omega t)} + \text{c.c.}$, where $f(r)$ denotes a general profile function. 
Since the bulk is incompressible at low-energy, the spatial profile of the probe is irrelevant deep in the bulk ($r\!\approx\!0$). A straightforward generalization of our calculations above~\cite{SI} yields $\Delta {S}\!=\! \sum_{n\in {\rm edge}} ({M}^{n,0}_+-{M}^{n,0}_-)\!=\!|f(R)|^2 |C_{\rm MB}|$:~While the edge response encodes the many-body Chern number, it is also sensitive to the probe intensity at the edge ($r\!=\!R$). In particular, one recovers $\Delta {S}\!=\!|C_{\rm MB}|A/\pi$ in the case of a uniform circular drive ($f(r)\!=\!r$).
This analysis indicates that it is possible to extract the many-body Chern number of correlated insulators {\it locally} from the edge, e.g.~employing a Laguerre-Gaussian-type probe~\cite{Goldman_LG_2012,Binanti_edge} with $f(r)\approx1$ at the system's edge and $f(r)\approx 0$ otherwise, which solely excites the boundary of the system~\cite{SI}.

{\it Concluding remarks---} Cold atoms in optical lattices~\cite{Cooper19_RMP} appear as a natural platform to measure the quantized edge response explored in this work:~the edge-mode configuration could be finely controlled by adjusting the confining potential~\cite{navon2021quantum}, and angular-momentum-sensitive spectroscopy could be applied locally on the edge of the cloud, using Laguerre-Gaussian beams~\cite{Goldman_LG_2012,Binanti_edge,lunt2024realization}. These methods could be directly applied to few-atom FCI states recently realized in experiments~\cite{leonard2023realization,lunt2024realization}. In these settings, the edge spectrum is highly discretized, such that the relevant matrix elements $M_{\pm}$ could be directly extracted from Rabi oscillations~\cite{Binanti_edge,TranCooperGoldman18_PRA_circdichroLL,nardin2024quantum}. It would be interesting to investigate extensions to non-Abelian topological matter, displaying  exotic edge structures~\cite{Pfaffian_atoms}.  Finally, understanding how the intrinsic geometric degrees of freedom of fractional QH states~\cite{Haldane_PRL_2013,Xavier_2025} influence dichroic rates -- both at the edge and in the bulk -- constitutes an intriguing avenue for further investigation.

{\it Acknowledgments ---}
The authors thank Iacopo Carusotto, Jean Dalibard, Daniele de Bernardis, Leonardo Mazza, Blagoje Oblak, Cecile Repellin and Peter Zoller for stimulating discussions.
F.N.\"U.~acknowledges funding from the Royal Society Grant No.~URF/R1/241667, the Marie Sk{\l}odowska-Curie programme of the European Commission No 893915, Simons Investigator Award [Grant No. 511029] and Trinity College Cambridge. 
A.~N.~thanks financial support from LoCoMacro 805252 from the European Research Council. 
N.~G.~is supported by the ERC Grant LATIS, the EOS project CHEQS and the Fondation ULB.

{\it Data Availability Statement---} The data is available from the authors upon reasonable request.

{\it Author Contribution Statement---} All authors contributed equally to this work.

\let\oldaddcontentsline\addcontentsline
\renewcommand{\addcontentsline}[3]{}

\let\addcontentsline\oldaddcontentsline

\clearpage
\cleardoublepage
\begin{supplementalMaterials}
\title{\color{Nathanblue}Supplemental Material for\\ ``Circular dichroism on the edge of quantum Hall systems:\\~From many-body Chern number to anisotropy measurements"}

\maketitle
{
	\hypersetup{linkcolor=black}
	\tableofcontents
}

\section{Quantized circular Dichroism:~excitation rates and connection to the Hall conductivity}
In this Appendix, we briefly review some of the key concepts at the basis of the circular dichroism.
We begin by introducing the conductivity tensor for quantum Hall systems.
We then review the connection between the power absorbed by a circularly polarized electric field and the conductivity tensor.
We highlight the connection with quantum mechanical rates. 
We will take advantage of the context to introduce in a more extensive manner the relevant quantities used in the main text and the notation employed therein.

The discussion here is mostly a re-elaboration of Refs.~\cite{Stern_Faraday,Tran17_SciAdv_dichrosim,goldman2024relating}.

\subsection{Conductivity tensor}
When an electric field $\mathbf{E}(\mathbf{r},t)$ is applied to a 2D material, within linear response, a current will flow proportional to the applied field. The current density  $J_i(\mathbf{r},t)$ can be written as
\begin{equation}
\label{eq:currentDensity}
J_i(\mathbf{r},t) = \int d^2\mathbf{r}' \int dt' \sum_j \sigma_{ij}(\mathbf{r}-\mathbf{r}', t-t') E_j(\mathbf{r}',t').
\end{equation}
$\sigma_{ij}$ is called the conductivity tensor, and under the assumption of isotropy and time-independence it can only depend on relative space-time displacements and must take the general form
\begin{equation}
\label{eq:conductivity_tensor}
\sigma =
\begin{pmatrix}
	\sigma_{xx} & \sigma_{xy}\\
	-\sigma_{xy} & \sigma_{xx}
\end{pmatrix}.
\end{equation}
We here introduce our Fourier conventions
\begin{equation}
\begin{split}
	\widetilde{E}_i(\mathbf{k},\omega)	&= \frac{1}{(2\pi)^3}\int d^2\mathbf{r} \int dt\,e^{-i (\mathbf{k}\cdot\mathbf{r}-\omega t)} \,E_i(\mathbf{r},t), \\
	E_i(\mathbf{r},t) &=\int d^2\mathbf{k} \int d\omega\,  e^{i (\mathbf{k}\cdot\mathbf{r}-\omega t)}\,\widetilde{E}_i(\mathbf{k},\omega),
\end{split}
\end{equation}
in terms of which one can write Eq.~\eqref{eq:currentDensity} as a completely local relation in frequency-momentum space
\begin{equation}
\widetilde{J}_i(\mathbf{k},\omega) =
(2\pi)^3\sum_j \widetilde{\sigma}_{ij}(\mathbf{k},\omega) \,\widetilde{E}_j(\mathbf{k},\omega).
\end{equation}

In the quantum Hall regime (in the bulk), at very long wavelengths, the current response is purely transverse and local, and quantized in terms of the many-body Chern number $C_{\rm MB}$. Namely,
\begin{equation}
\sigma_{ij}(\mathbf{r}-\mathbf{r}', t-t') = - \frac{q^2}{h}C_{\rm MB}\,\epsilon_{ij}\,\, \delta(t-t')\delta(\mathbf{r}-\mathbf{r}')	,
\end{equation} 
with $\epsilon_{ij}$ a totally antisymmetric tensor and $q$ denotes the charge of the carriers.
From Eq.~\eqref{eq:currentDensity}, one recovers the familiar quantized transverse response
\begin{equation}
J_i(\mathbf{r},t) = - \frac{q^2}{h}C_{\rm MB}\,\sum_j\epsilon_{ij} E_j(\mathbf{r},t).
\end{equation}
The Fourier transform of the conductivity tensor therefore reads, with the conventions we adopted,
\begin{equation}
\label{eq:HallResponse}
\begin{split}
	\widetilde{\sigma}_{ij}(\mathbf{k},\omega)	&= - \frac{1}{(2\pi)^3} \,\frac{q^2}{h}C_{\rm MB}\,\,\epsilon_{ij}.
\end{split}
\end{equation}

\subsection{Circular dichroism}
We here consider a monochromatic, uniform and circularly polarized electric field $\mathbf{E}=\mathbf{E}_\pm$ given by
\begin{equation}
\label{eq:electric_field}
\mathbf{E}_{\pm}(\mathbf{r},t) = 2E e^{- i \omega t} \,(\hat{e}_x\pm i \hat{e}_y),
\end{equation}
impinging on a 2D material in the $x-y$ plane, of area $A$.
For convenience, we use the complex electric field framework~\cite{Stern_Faraday}; the physical electric field is the real part of $\mathbf{E}_{\pm}$, and can be seen to equal $E e^{- i \omega t} \,(\hat{e}_x\pm i \hat{e}_y)+\text{c.c.}$, i.e.~the same electric field that we introduced in the main text.
For notational convenience we also introduce
\begin{equation}
\hat{\epsilon}_{\pm} = \hat{e}_x\pm i \hat{e}_y.
\end{equation}

The electric field of Eq.~\eqref{eq:electric_field} enters the Schr\"odinger equation through an additional scalar potential of the form $U_{\pm}(\mathbf{r},t) = -2qE (x \cos(\omega t) \pm y \sin(\omega t))$, or, equivalently, in the form of an additional vector potential $\mathbf{A}_{\pm}(\mathbf{r},t)=\frac{2E}{\omega}(\sin(\omega t) \hat e_x \mp \cos(\omega t) \hat e_y)$. In this article, we generally adopt  the first approach. We note that such a circular drive can also be implemented in cold-atom setups through a circular shaking of the optical lattice~\cite{Asteria19_NatPhys}.

The time-averaged (over a period $2\pi/\omega$) power absorbed by the material,
\begin{equation}
P_\pm(\omega) = \int d^2\mathbf{r} \, \overline{\Re \mathbf{J}\cdot \Re \mathbf{E}_{\pm}} =
\frac{1}{2}\Re \int d^2\mathbf{r} \, \mathbf{J} \cdot \mathbf{E}_{\pm}^*,
\end{equation}
can be related to the conductivity tensor Eq.~\eqref{eq:conductivity_tensor} by using the current density expression Eq.~\eqref{eq:currentDensity}; after few algebraic manipulations one obtains
\begin{equation}
\label{eq:absorbed_power}
P_\pm(\omega) =	(2\pi)^3\,4 E^2 A \left(\widetilde{\sigma}_{xx}^R(\omega) \mp   \widetilde{\sigma}_{xy}^I(\omega)\right).
\end{equation}
Here, $\widetilde{\sigma}_{ij}(\omega) = \widetilde{\sigma}_{ij}(\mathbf{k}=0, \omega)$ and the $R(I)$ superscript indicate the real(imaginary) part.
Importantly the transverse part of the conductivity tensor, $\sigma_{xy}$, is related to circular dichroism:~the power absorbed $P_{\pm}$ associated with opposite circular polarizations $\hat{\epsilon}_{\pm}$ can be different. By taking the difference one indeed obtains a non-zero result,
\begin{equation}
\label{eq:circular_dichroism}
\Delta P(\omega) = -8(2\pi)^3\,E^2 A \, \widetilde{\sigma}^I_{xy}(\mathbf{k}=0,\omega).
\end{equation}
In an isotropic system $\sigma_{ij}(\mathbf{r},t)=\sigma_{ij}(-\mathbf{r},t)$ the real and imaginary parts of the conductivity tensor can be related by the famous Kramers-Kronig realations,
\begin{equation}
\widetilde{\sigma}_{xy}^R(\omega) = \frac{2}{\pi} \dashint_{0}^{\infty}  \frac{\omega'\,\widetilde{\sigma}_{xy}^I(\omega')}{(\omega')^2-\omega^2} \,d\omega',			
\end{equation}
which, in the $\omega\rightarrow0$ limit, yield a sum-rule through which one relates the circular dichroism in Eq.~\eqref{eq:circular_dichroism} to the Hall transverse response,
\begin{equation}
\label{eq:dichroism_HallResponse}
\frac{2}{\pi} \dashint_0^{\infty} \frac{\Delta P(\omega)}{\omega}\,d\omega = -8
(2\pi)^3\,E^2 A \, \widetilde{\sigma}_{xy}^R(0).
\end{equation}	

\subsection{Connection to quantum mechanical rates}
The absorbed power can be related to quantum-mechanical rates: Since the absorbed power at the given frequency is the energy absorbed per unit time, and the energy is absorbed in quanta $\hbar\omega$, we can write
\begin{equation}
\frac{P_\pm(\omega)}{\hbar\omega} = \Gamma_\pm(\omega),
\end{equation}
where $\Gamma_\pm(\omega)$ is the transition rate from the system's ground state $\ket{0}$ to any other state. 
We point out that the rate difference $\Delta\Gamma(\omega) = \Gamma_+(\omega)-\Gamma_-(\omega)$ is the quantity that appears in the left hand side of Eq.~\eqref{eq:dichroism_HallResponse}.

The rates can be computed within the scope of perturbation theory. One can indeed see the impinging electric field Eq.~\eqref{eq:electric_field} as a time-dependent external potential,
\begin{equation}
\label{eq:dichroic_electric_field_potential_energy}
{U}_{\pm}(\mathbf{r},t) = \Re \left[ -2q E e^{- i \omega t} \,(x\pm i y)\right],
\end{equation}
being applied to the system.
Using Fermi's golden rule one obtains
\begin{equation}
\label{eq:rates}
\Gamma_{\pm}(\omega>0) = 4 \pi\,\frac{q^2E^2}{\hbar^2} \sum_n M_{\pm}^{n,0}  \delta(\omega-\omega_{n,0}).
\end{equation}
Here, $\omega_{n,0} = \frac{\varepsilon_n-\varepsilon_0}{\hbar} \geq0$ are the (many-body) Bohr's frequencies and the matrix elements $M_\pm^{n,0}$ are the same as those we introduced in the main text in Eq.~{\color{red!95!black}5}, namely
\begin{equation}
\label{eq:CD_matrix_elements}
M_{\pm}^{n,0} = \left|\int (x\pm i y)\Braket{n| \hat\rho(\mathbf{r})|0} d^2\mathbf{r}\right|^2,
\end{equation}
with $\hat\rho(\mathbf{r})=\sum_{j=1}^N \delta^{(2)}(\mathbf{r}_i-\mathbf{r}_j)$ being the particle density operator.

When, in the spirit of Eq.~\eqref{eq:dichroism_HallResponse}, the rates are integrated over all the (positive) frequencies, one obtains
\begin{align}
\label{eq:integrated_rates}
\Gamma_\pm^{\rm int}&=\int_{0}^\infty \Delta\Gamma(\omega)d\omega=4 \pi\,\frac{q^2E^2}{\hbar^2} S_{\pm},\\
S_{\pm}&=\sum_n M_{\pm}^{n,0}.
\end{align}
These equations link the frequency integrated transition rates $\Gamma_\pm^{\rm int}$ to the summation of all the possible transition matrix-elements, $S_{\pm}=\sum_n M_{\pm}^{n,0}$.

Using this result in conjunction with Eq.~\eqref{eq:dichroism_HallResponse}, we can write
\begin{equation}
\label{eq:rates_difference}
\frac{\hbar^2}{2q^2E^2A} \,\left(\Gamma_+^{\rm int}-\Gamma_-^{\rm int}\right)=-(2\pi)^3\frac{\widetilde{\sigma}_{xy}^R(0)}{q^2/h}.
\end{equation}
If the system is infinite (pure bulk), then $\widetilde{\sigma}_{xy}^R(0)$ is given by Eq.~\eqref{eq:HallResponse}, yielding
\begin{equation}
\label{eq:bulk_dichroism}
\frac{\hbar^2}{2q^2E^2A} \,\left(\Gamma_+^{\rm int}-\Gamma_-^{\rm int}\right)_{\rm bulk}=+C_{\rm MB},
\end{equation}
which is the same formula we quoted in the main text [see Eq.~{\color{red!95!black}2}], when setting the charge $q\!=\!1$.

Importantly, if the system is finite, then no net current can be present at any time. Namely, it must hold true that
\begin{equation}
\int J_i(\mathbf{r},t) d^2\mathbf{r} = 0.
\end{equation}
This, together with Eq.~\eqref{eq:conductivity_tensor}, implies that $\int \sigma_{ij}(\mathbf{r}-\mathbf{r}, t)  d^2\mathbf{r} = 0$, since the applied electric field is generic.
Equivalently,
\begin{equation}
\widetilde{\sigma}_{ij}(\mathbf{k}=0,\omega)=0,
\end{equation}
which implies (see Eq.~\eqref{eq:rates_difference}) that
\begin{equation}
\label{eq:cancellation}
\frac{\hbar^2}{2q^2E^2A} \,\left(\Gamma_+^{\rm int}-\Gamma_-^{\rm int}\right)=0.
\end{equation}
This means that the frequency integrated transition rates for a clockwise and anticlocwise polarized electric field exactly cancel out.

Expressed differently, one can state that the bulk dichroic response given by Eq.~\eqref{eq:bulk_dichroism} must be exactly  canceled by a second contribution, which must originate from the presence of an edge in a finite system; namely,
\begin{equation} 
\left( \Gamma^{\text{int}}_+ - \Gamma^{\text{int}}_-  \right)_{\text{bulk}} = - \left( \Gamma^{\text{int}}_+ - \Gamma^{\text{int}}_-  \right)_{\text{edge}},
\end{equation}
as we discussed in the main text; see Eq.~{\color{red!95!black}3}.

\section{Quantized circular-dichroic response of edge modes}
In this Appendix, we provide a more detailed derivation of the low-energy dichroic response, obtained from the viewpoint of the \cll theory, which is sketched in the main text. We also take this opportunity to present two natural extensions, considering situations involving multiple edge-mode branches, and QH droplets with non-circular geometry.

First, we will assume that the fractional quantum Hall liquid has an approximately circular shape, with constant density throughout \mbox{$\rho_0=\nu/(2\pi l_B^2)$}, occupying an area $\pi R^2$, with \mbox{$R=\sqrt{2N/\nu}l_B$}.
Here $\nu=-C_{\rm MB}$ is the (fractionally) quantized filling fraction of the system and $l_B=\sqrt{\hbar/qB}$ the magnetic length.

In the following, we will thus be computing $S_\pm$ [see Eq.~{\color{red!95!black}5}] from a low-energy perspective.
For the sake of generality, we will be considering a perturbing potential of the form $U_\pm(\mathbf{r},t) = -qE\,(f(r) e^{i(\pm\theta-\omega t)} + \text{c.c.})$ which has an arbitrary radial profile $f(r)$ but still carries one unit of angular momentum.
The relevant matrix elements [compare to Eq.~\eqref{eq:integrated_rates}] then read
\begin{align}
\label{eq:S}
S_{\pm}&=\sum_{n\in\text{edge}} \left|v_\pm^{n,0}\right|^2,
\\
\label{eq:matrix_elements}
v_\pm^{n,0} &=  \int   f(r)e^{\pm i \theta} \Braket{n\left| \hat{\rho}(\mathbf{r}) \right|0} d{\bf r},
\end{align}
which generalize those presented in Eq.~{\color{red!95!black}5} for the more standard circular-dichroic setting (i.e.~a uniform circularly-polarized electric field).
Contrary to Eq.~\eqref{eq:integrated_rates}, here the summation is taken over the low-energy edge-modes only, so as to isolate the edge contribution from the bulk one.

Since the bulk is incompressible, contributions to the integral appearing in Eq.~\eqref{eq:matrix_elements}
can only come from the edge region, located at $r\simeq R$, with a typical width set by the magnetic length $l_B$.
As a consequence, one can approximate
\begin{equation}
\label{eq:step1}
v_\pm^{n,0} = f(R)\, \int d\theta \,e^{\pm i \theta}\Braket{n| \hat{\rho}_{\rm eff}(\theta) |0},
\end{equation}
where
\begin{equation}
\label{eq:effective_1d_density}
\hat{\rho}_{\rm eff}(\theta) = \int {\rho}(\mathbf{r}) \,r\,dr
\end{equation}
is a one-dimensional (edge) density operator obtained by integrating out the direction orthogonal to the edge.
As $R\rightarrow\infty$, i.e. when the number of particles $N$ is large, the approximation becomes more and more accurate provided $f(r)$  varies slowly over the edge width $\propto l_B$.

We now recognize the $\ell=\pm1$ Fourier components $$\hat{\rho}_{\ell}=\int d\theta \,e^{i\ell \theta} \hat{\rho}_{\rm eff}(\theta),$$ of the edge density operator, and rewrite Eq.~\eqref{eq:step1} as
\begin{equation}
\label{eq:step2}
v_\pm^{n,0} = f(R)\, \Braket{n | \hat{\rho}_{\pm1} | 0}.
\end{equation}

In the following, we will first discuss the simplest case of a single edge-mode, 
and then discuss the case of multiple edges.

\subsection{The case of a single circular edge}
Following Wen~\cite{Wen_AdvPhys_1995,Hansson_RMP_2017}, 
the excess density along the edge of a QH droplet fails to commute, and leads to a $U(1)$ Kac-Moody algebra, 
\begin{equation}
\label{eq:xll_density_commutators}
[\hat{\rho}_{\ell},\hat{\rho}_{\ell'}]=\nu \ell \delta_{\ell,-\ell'}. 
\end{equation}
As previously, $\nu$ represents the (fractionally quantized) filling fraction defined in the bulk of the QH droplet, and it naturally comes into play through the average bulk density, $\rho_0=\nu/(2 \pi l_B^2)$.

As a consequence of Eq.~\eqref{eq:xll_density_commutators}, the operators $\hat{\rho}_{\ell}$ can be identified as bosonic annihilation (/creation) operators
\begin{align}
\label{eq:single_edge_bosonization}
\begin{cases}
	\hat{\rho}_{\ell>0} = \sqrt{\nu|\ell|} \, \hat{b}_\ell^\dagger
	\\
	\hat{\rho}_{\ell<0} = \sqrt{\nu|\ell|} \, \hat{b}_\ell^\nodagger,
\end{cases}
\\
\label{eq:commutation_relations}
[\hat{b}_\ell^\nodagger,\hat{b}_\ell^\dagger]=\delta_{\ell,\ell'}.
\end{align}
The ground state $\ket{0}$ is identified as the vacuum of bosonic excitations, while the excited states are obtained by acting on it with the creation operators $\hat{b}_\ell^\dagger$.

It follows immediately that, for every possible edge state $\ket{n}$, the matrix elements $v_-^{n,0}=0$ entering Eq.~\eqref{eq:step2} must vanish. 
The ground state $\ket{0}$ is indeed annihilated by $\hat{\rho}_{-1}$; it follows that  $S_-=0$.
This can be easily seen as a consequence of the presence of a single chiral edge mode: $\hat{\rho}_{-1}$ aims at lowering the ground state angular momentum, but there is no gapless edge mode (with the corresponding angular momentum) that can be excited.

Therefore, in the circularly symmetric case with a single chiral branch of edge modes, only $v_+^{n,0}$ can be non-zero.
By inspection of Eqs.~\eqref{eq:step2}-\eqref{eq:single_edge_bosonization} it is straightforward to realize that the only state contributing to the sum in Eq.~\eqref{eq:S} is $\ket{1}=\hat{b}_{1}^\dagger\ket{0}$, and thus
\begin{equation}
S_{+} = v_+^{1,0} = |f(R)|^2 \nu.
\end{equation}
By identifying the quantized filling fraction $\nu$ with the many-body Chern number, $\nu=-C_{\rm MB}$, the previous equation can equivalently be written as
\begin{equation}
S_{+}-S_{-} = S_{+} = - |f(R)|^2 C_{\rm MB} = |f(R)|^2 |C_{\rm MB}|,
\end{equation}
where we assumed $C_{\rm MB}\!<\!0$ in the last step; see also the summary below.

In the conventional circular-dichroism case (uniform circular drive), $f(R)\!=\!R$ and the result can be rewritten as
\begin{equation}
\label{eq:edge_dichro}
S_{+}-S_{-} =  \frac{A}{\pi} |C_{\rm MB}|,
\end{equation}
which is the result quoted in the main text.

To summarize, we obtained that the ``ideal" case of a single (circular) edge mode is characterized by the responses
\begin{align}
&S_+ = - C_{\rm MB} \text{ and } S_- =0 \text{ on the edge}, \notag\\
&S_- = - C_{\rm MB} \text{ and } S_+ =0 \text{ in the bulk}, \notag
\end{align}
where we invoked Eq.~({\color{red!95!black}3}). As previously noted, we emphasize that these results correspond to a situation where the many-body Chern number is $C_{\rm MB}\!<\!0$. 

We note that the case $C_{\rm MB}\!>\!0$ is directly related to the previous situation by a time-reversal transformation, i.e.~by replacing $C_{\rm MB}\rightarrow - C_{\rm MB}$ and $S_{\pm}\rightarrow S_{\mp}$. It follows that for  $C_{\rm MB}>0$, the ``ideal" case is characterized by $S_- \!=\! C_{\rm MB}$ and $S_+ \!=\!0$ on the edge, and correspondingly $S_+ \!=\! C_{\rm MB}$ and $S_- \!=\!0$ in the bulk.

\subsection{The case of multiple circular edges}
Considering the case of multiple (spatially separated) edges, potentially displaying modes of different nature, $S_-$ can now be finite. Indeed, in this case, the action of $\hat{\rho}_{\ell<0}$ on the ground state can be non-trivial and contribute to the total excitation rate.

When multiple edge modes are present, the effective one-dimensional edge density $\hat{\rho}_{\rm eff}(\theta)$ can be split into several density operators, each associated with a different edge channel,
\begin{equation}
\label{eq:edge_splitting}
\hat{\rho}_{\rm eff}(\theta) = \sum_{e} \hat{\rho}_e(\theta).
\end{equation}
Here, the sum runs over all the edge modes 
and $\hat{\rho}_e(\theta)$ is the density operator associated to a given channel.
It is useful to introduce the Fourier transform of each component $\hat{\rho}_e(\theta)$ as
\begin{equation}
\hat{\rho}_{e,\ell}=\int d\theta \,e^{i\ell \theta} \hat{\rho}_{e}(\theta).
\end{equation}
Analogously to Eq.~\eqref{eq:single_edge_bosonization}, these operators fail to commute. 
When the edges are well separated (by a distance much larger than the magnetic length $l_B$), they satisfy~\cite{Wen_AdvPhys_1995}
\begin{equation}
\label{eq:multiple_edges_commutations}
[\hat{\rho}_{e,\ell},\hat{\rho}_{e',\ell'}] = \nu_e\,\delta_{e,e'}\delta_{\ell,-\ell'}.
\end{equation}
Here $\nu_e$ is the (signed) difference between the filling fractions of the fluids sitting on the opposite sides of the edge, the vacuum corresponding to zero filling.
As a consequence, these numbers fulfill the constraint $\sum_e \nu_e=\nu$, where $\nu$ is the bulk filling fraction, i.e.~(minus) the many-body Chern number.

As a consequence of Eq.~\eqref{eq:multiple_edges_commutations}, provided $\nu_e>0$, one can introduce bosonic operators as
\begin{equation}
\begin{cases}
	\hat{\rho}_{e,\ell>0} &= \sqrt{\nu_e|\ell|}\,\hat{b}_{e,\ell}^\dagger,
	\\
	\hat{\rho}_{e,\ell<0} &= \sqrt{\nu_e|\ell|}\,\hat{b}_{e,\ell}^\nodagger.
\end{cases}
\end{equation}
In the other case, i.e.~when $\nu_e<0$, we instead have
\begin{equation}
\begin{cases}
	\hat{\rho}_{e,\ell>0} &= \sqrt{-\nu_e|\ell|}\,\hat{b}_{e,\ell}^\nodagger,
	\\
	\hat{\rho}_{e,\ell<0} &= \sqrt{-\nu_e|\ell|}\,\hat{b}_{e,\ell}^\dagger.
\end{cases}
\end{equation}
Using these identifications together with Eq.~\eqref{eq:edge_splitting}, one can thus rewrite the $\ell=\pm1$ Fourier components of $\hat{\rho}_{\rm eff}(\theta)$ as
\begin{equation}
\begin{split}
	\hat{\rho}_{1} &= \sum_{e, \nu_e>0} \sqrt{\nu_e} \hat{b}_{e,1}^\dagger + \sum_{e, \nu_e<0} \sqrt{-\nu_e} \hat{b}_{e,1}^\nodagger,
	\\
	\hat{\rho}_{-1} &= \sum_{e, \nu_e>0} \sqrt{\nu_e} \hat{b}_{e,1}^\nodagger + \sum_{e, \nu_e<0} \sqrt{-\nu_e} \hat{b}_{e,1}^\dagger.
\end{split}
\end{equation}

Analogously to the single-edge case, only a single state (for every edge mode $e$) $\ket{e}={b}_{e,1}^\dagger\ket{0}$ contributes to the relevant matrix elements $v_{\pm}^{n,0}$ in Eq.~\eqref{eq:step1}.
Namely,
\begin{equation}
\begin{split}
	v_+^{e,0} &= f(R)\, \sqrt{\nu_e}, 
	\\
	v_-^{e,0} &= f(R)\, \sqrt{-\nu_e}.
\end{split}
\end{equation}
The contributions from the two chiralities $S_\pm$ [see Eq.~\eqref{eq:S}] can therefore be evaluated; the results read
\begin{equation}
\begin{split}
	S_+ &= |f(R)|^2 \sum_{e,\nu_e>0} \nu_e,
	\\
	S_- &= |f(R)|^2 \sum_{e,\nu_e<0} -\nu_e.
\end{split}
\end{equation}
Here, we note that none of these individual contributions is robustly quantized in terms of the many-body Chern number. 
However, the difference $S_+-S_-$ is quantized, since
\begin{equation}
\begin{split}
	\Delta S&=S_+-S_-\\ 
	&= |f(R)|^2 \left(\sum_{e,\nu_e>0} \nu_e + \sum_{e,\nu_e<0} \nu_e\right) 
	\\
	&= |f(R)|^2 \sum_{e} \nu_e = |f(R)|^2 \nu,
\end{split}
\end{equation}
which is precisely the result quoted in the main text.

In the case of conventional circular dichroism (uniform circular driving), one sets $f(R)\!=\!R$ and the result can be written as
\begin{equation}
\label{eq:delta_S_circular_dichroism}
\Delta S = -\frac{A}{\pi} C_{\rm MB} = \frac{A}{\pi} |C_{\rm MB}|;
\end{equation}
as before, the last equality is valid when $C_{\rm MB}<0$.

We conclude this section with a series of comments and remarks. Here, we explicitly assumed that: (i) the various edges are sufficiently close to each other, such that they are all located at $r\simeq R$; and (ii) we assumed that these edges are well separated one from the other. 
The first assumption (i) should not be harmful in the thermodynamic limit, provided the confinement is steep, since the particle density can be expected to suddenly drop from its bulk value $\rho_0=\frac{\nu}{2\pi l_B^2}$ to zero.
However it must be noticed that this is not always the case, especially in the case of smooth confining potentials, where ``wedding-cakes" of different filling fractions can be expected~\cite{Cooper_PRA_2005}.
On the other hand, we do not expect the second assumption (ii) to be crucial, provided that when the edges are brought together there exists a unitary mapping between the two sets of modes (i.e. those associated to the ``far-away" edges and the ones of the ``close-by" case).

\subsection{The case of non-circular boundaries}
We finally demonstrate that the total circular dichroic response in Eq.~\eqref{eq:edge_dichro} does not rely on the boundary being circular. Specifically, we will now consider QH droplets whose shape is determined by some externally imposed confinement potential, of arbitrary shape~\cite{Oblak_PRX_2024}. 
In this case, the boundary of the QH system will follow one of the equipotential lines of the confinement, provided the QH droplet is large enough ($N\gg1$) and the equipotential does not have ``too sharp" features (on the scale set by the magnetic length $l_B$).
For the sake of simplicity, we hereby consider the case of a single chiral edge mode propagating along this non-trivial geometry.

We consider a convenient set of (action-angle~\cite{Oblak_PRX_2024}) coordinates $(K,\Theta)$ in terms of which we express the Cartesian coordinates
\begin{equation}
\label{eq:coordinates}
\begin{cases}
	x = F(K,\Theta)
	\\
	y = G(K,\Theta).
\end{cases}
\end{equation}
Curves at constant $K$ describe, at least locally at the position of the system's boundary, equipotential lines for the confinement, i.e.~they follow the system's boundary. 
Notice that the angle variable $\Theta$ does not necessarily correspond to the polar angle~\cite{Oblak_PRX_2024}.

The boundary will thus be parameterized by a curve described by a specific value of $K$, say $K=K_{\rm edge}$. We decompose the parametric representation of the boundary in Fourier modes as
\begin{equation}
\label{eq:x_y_at_the_edge}
\begin{cases}
	x = F(K_{\rm edge},\Theta)=\sum_\ell e^{i \ell \Theta}f_l,
	\\
	y = G(K_{\rm edge},\Theta)=\sum_\ell e^{i \ell \Theta}g_l.
\end{cases}
\end{equation}
Here, since $F(K_{\rm edge},\Theta)$ and $G(K_{\rm edge},\Theta)$ are real valued, $f_{-\ell}=f_\ell^*$ and $g_{-\ell}=g_\ell^*$.

We now compute the relevant transition matrix elements [see Eq.~\eqref{eq:CD_matrix_elements}],
\begin{equation}
v_\pm^{n,0} =  \int  (x\pm i y) \Braket{n\left| \hat{\rho}(\mathbf{r}) \right|0} d{\bf r}.
\end{equation}
On the system's boundary, one can approximate $x\pm i y$ using Eq.~\eqref{eq:x_y_at_the_edge} and obtain
\begin{equation}
v_\pm^{n,0} =  \sum_\ell (f_\ell\pm i g_\ell) \int e^{i \ell \Theta} \Braket{n\left|\hat{\rho}_{\rm eff}(\Theta)\right|0} d{\Theta},
\end{equation}
where, in analogy with Eq.~\eqref{eq:effective_1d_density}, we identify the $K$ integral of the two-dimensional particle density operator $\hat{\rho}(\mathbf{r})$, 
\begin{equation}
\hat{\rho}_{\rm eff}(\Theta) = \int  \hat{\rho}\bigl(x(K,\Theta),y(K,\Theta)\bigr) dK ,
\end{equation}
as the effective one-dimensional density operator of the \cll theory. Therefore, (analogously to Eq.~\eqref{eq:single_edge_bosonization} and Eq.~\eqref{eq:commutation_relations}), 
the Fourier components $\hat{\rho}_\ell=\int e^{i \ell \Theta} \hat{\rho}_{\rm eff}(\Theta)d\Theta$ can be identified with bosonic creation/annihilation operators, and the excited edge-states $\ket{n}$ as states in which \cll quanta have been inserted by acting with bosonic creation operators on the vacuum state $\ket{0}$. 
With these identifications, we obtain
\begin{equation}
v_\pm^{n,0} =  \sqrt{\nu n} (f_n\pm i g_n),
\end{equation}
and thus
\begin{equation}
\label{eq:Spm_deformed}
S_\pm = |C_{\rm MB}|\,\sum_{n>0} n \left|f_n\pm i g_n\right|^2.
\end{equation}
Here we have again written $|C_{\rm MB}|=\nu$, since we have $C_{\rm MB}<0$.
Importantly, in contrast with the circular-geometry case, $S_-$ does not necessarily vanish in the presence of an arbitrary edge geometry, even when the boundary features a single chiral mode.

After simple algebraic manipulations, the differential rate $\Delta S=S_+-S_-$ can finally be written as
\begin{equation}
\label{eq:deltas_preliminary}
\Delta S = |C_{\rm MB}|\,\frac{1}{\pi}\left(-2\pi i\sum_{n>0} n (f_n g_n^* - f_n^* g_n)\right).
\end{equation}

We notice that the area of the quantum Hall system coincides with the area enclosed by the parametric curve in Eq.~\eqref{eq:x_y_at_the_edge} and can be written as
\begin{equation}
A=\frac{1}{2}\oint (x dy- y dx) = -2\pi i\sum_{\ell>0} \ell \left(f_\ell g_\ell^* - f_\ell^* g_\ell \right),
\end{equation}
which is the same expression appearing on the right hand side of Eq.~\eqref{eq:deltas_preliminary}. We therefore recover the general result,
\begin{equation}
\label{eq:dichroism_arbitrary_shape}
S_+ - S_- =- \frac{A}{\pi}\,C_{\rm MB}.
\end{equation}
This demonstrates that the edge-dichroic response is entirely low-energy in nature and it is quantized, even for general boundary shapes. In this case, both $S_\pm$ contribute to the response, and their specific values are dictated by the shape of the system's boundary. However, their difference depends on the geometry only through the system's area, $A$. In fact, and as we further discuss below, their ratio $S_-/S_+$ measures the anisotropy of the boundary.

As a concrete example, we consider the anisotropic droplets described in Ref.~\cite{Oblak_PRX_2024}.
In particular, we consider a confinement potential given by
\begin{equation}
\label{eq:confinement_deformed}
V_{k,\lambda}(r,\theta)=\frac{u_0}{2}r^2(\cosh(2\lambda)+\cos(k\theta)\sinh(2\lambda)),
\end{equation}
with polar coordinates $r, \theta$, and $k$ is an integer, while $u_0$ and $\lambda$ set the scale and deformation's magnitude of the confinement, respectively. 
We point out that any sufficiently smooth confinement potential that depends only on $r^2(\cosh(2\lambda)+\cos(k\theta)\sinh(2\lambda))$ exhibits the same properties~\cite{Oblak_PRX_2024}. This observation also includes, in particular, the case of anharmonic--anisotropic traps.
Furthermore, the specific analytical form of the confinement deep in the incompressible bulk, is irrelevant for the results:~all that matters, indeed, is the shape of the edge.

In terms of the angle coordinate $\Theta$, the polar angle $\theta$ reads
\begin{equation}
\theta(\Theta)=\frac{1}{k}\arg\left(\frac{\sinh(\lambda)-e^{ik\Theta}\cosh(\lambda)}{-\cosh(\lambda)+e^{ik\Theta}\sinh(\lambda)}\right)+2\pi\frac{n}{k},
\end{equation}
where $n=0,1,\hdots k-1$.
The shape of the boundary in Eq.~\eqref{eq:x_y_at_the_edge} can then be expressed as
\begin{equation}
\label{eq:boundary_shape}
\begin{cases}
	\frac{x/l_B}{\sqrt{2K_{\rm edge}}}\!=\!\sqrt{\cosh(2\lambda)-\cos(k\Theta)\sinh(2\lambda)}\,\cos(\theta(\Theta)),
	\\
	\frac{y/l_B}{\sqrt{2K_{\rm edge}}}\!=\!\sqrt{\cosh(2\lambda)-\cos(k\Theta)\sinh(2\lambda)}\,\sin(\theta(\Theta)),
\end{cases}
\end{equation}
which allows for the (at least numerical) evaluation of $S_{\pm}$ as given in Eq.~\eqref{eq:Spm_deformed}.

We notice that, in the present case, $K_{\rm edge}$ only appears as a scale factor for the boundary shape. As a consequence, the only dependence of $f_l$ and $g_l$ on $K_{\rm edge}$ is of the form $\sqrt{K_{\rm edge}}$ and thus in the thermodynamic limit (when the low-energy edge description holds) the ratio $S_{-}/S_{+}$ depends only on the geometry of the boundary.\\

The $k=2$ case is particularly simple. In this case, the potential in Eq.~\eqref{eq:confinement_deformed} has a simple quadratic form,
\begin{equation}
\label{eq:elliptic_potential}
V_{2,\lambda}(r,\theta)=\frac{u_0}{2}(e^{2\lambda}x^2+e^{-2\lambda}y^2).
\end{equation}
This deformed potential is relevant for rapidly rotating atomic gases~\cite{Fletcher_Science_2021}.

The boundary shape in Eq.~\eqref{eq:boundary_shape} simplifies to
\begin{equation}
\label{eq:boundary_shape_ellipse}
\begin{cases}
	x/l_B=\sqrt{2K_{\rm edge}}\,e^{-\lambda} \cos(\theta),
	\\
	y/l_B=\sqrt{2K_{\rm edge}}\,e^{+\lambda} \sin(\theta),
\end{cases}
\end{equation}
which is an ellipse with an enclosed area $A=2\pi K_{\rm edge}\,l_B^2$.
Furthermore, from Eq.~\eqref{eq:Spm_deformed}, we obtain
\begin{equation}
\label{eq:Spm_elliptic}
\begin{split}
	S_+ &= |C_{\rm MB}|\,2K_{\rm edge}\,l_B^2 \cosh^2(\lambda),
	\\
	S_- &= |C_{\rm MB}|\,\,2K_{\rm edge}\,l_B^2 \sinh^2(\lambda),
\end{split}
\end{equation}
which, as expected on the more general grounds exposed above, satisfies $\Delta S = -C_{\rm MB}\,2K_{\rm edge}l_B^2=-C_{\rm MB}\,A/\pi$.
Finally, the ratio
\begin{equation}
\frac{S_-}{S_+} = \tanh^2(\lambda),
\end{equation}
is indeed, as anticipated above on more general grounds, independent of $K_{\rm edge}$, and thus also independent of the number of particles; 
rather, it reflects the boundary ellipticity.

\section{Numerical calculations}
This Appendix provides more details on the numerical calculations presented in the main text, as well as additional numerical results, not included there, which corroborate our analysis but are not essential for understanding the work.

\subsection{Single-edge, circular boundary configurations}
We begin by presenting a more extensive discussion on our numerical results concerning the case of a QH system with a single edge mode when the boundary is (at least approximately) circular.
In particular, we first discuss more extensively our CI results based on the Harper-Hofstadter model; we also demonstrate, considering a single edge-mode for simplicity, how locally addressing the boundary of the system allows to meaningfully extract the bulk's Chern number.
We then introduce the FCI model (Bose-Hubbard Harper-Hofstadter) we considered and show analogous numerical results in a strongly-correlated scenario.
We finally demonstrate how the same physics can be observed in a different setup modeled by the Haldane honeycomb model, a prototypical example of a CI in the absence of a net magnetic field.

\subsubsection{Harper-Hofstadter CI}
We consider in this case the Harper-Hofstadter model~\cite{Cooper19_RMP}. 
Its Hamiltonian, on a square lattice and in the Landau gauge, reads
\begin{align}  \label{eq:Hofst.Hamiltonian}
H_0 &= -J \sum_{m,n} \left( e^{i2\pi\phi m}\,a^{\dagger}_{m,n+1} a^{\phantom{\dagger}}_{m,n} + a^{\dagger}_{m+1,n} a^{\phantom{\dagger}}_{m,n} +\text{h.c.} \right) \nonumber\\
& +\sum_{m,n} V_{m,n} \, a^{\dagger}_{m,n} a^{\phantom{\dagger}}_{m,n}.
\end{align}
It describes particles hopping on a lattice with nearest-neighbor tunneling amplitude $J$, under the influence of a uniform magnetic field with a flux $\phi$ per plaquette: a particle hopping around a plaquette picks up an Aharonov-Bohm phase $\exp(i2\pi\phi)$.
Here, $a^{(\dagger)}_{m,n}$ annihilates (creates) a particle at the lattice site $\br_{m,n}=m\hat{e}_x+n\hat{e}_y$. 
We set the lattice constant to unity, and $(m,n)\in\mathbb{Z}$.
We also include an additional confining potential $V_{m,n}$ which we use as a knob to control the edge-mode configuration. 
We note that trapping potentials can now be finely designed in ultracold atoms, using digital-micromirror or spatial-light-modulator techniques~\cite{navon2021quantum}.

\begin{figure}
\includegraphics[width=1\linewidth]{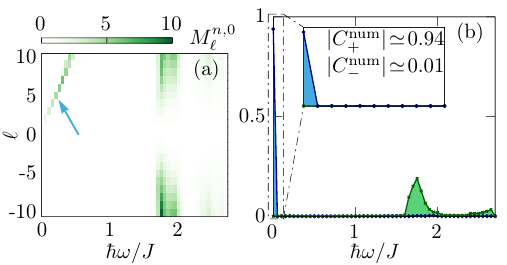}
\caption{
	(a) Angular-momentum-resolved spectroscopy in
	the single-particle Harper-Hofstadter model:~Matrix elements $M_{\ell}^{n,0}\!=\!\left|\int d^2\mathbf{r}\,  e^{i \ell \theta}\braket{n|\hat{\rho}(\mathbf{r})|0}\right|^2$, as a function of frequency, $\omega_{n,0}\!=\!(\varepsilon_n-\varepsilon_0)/\hbar$, and transferred (optical) angular momentum $\ell$; $\varepsilon_n$ is the energy of the $n$-th eigenstate of the single-particle Harper-Hofstadter model Eq.~\eqref{eq:Hofst.Hamiltonian}.
	The presence of a single edge mode is highlighted by the arrows.
	(b) CD matrix elements $M_{\pm}^{n,0}$ [Eq.~{\color{red!95!black}5}], as a function of $\omega_{n,0}$, with the inset zooming in on the low-energy features. The numerically extracted Chern number contributions, $C_{\text{MB}}^{\rm num} = C_+^{\rm num}-C_-^{\rm num}$, are calculated by isolating the low-energy response ($\hbar\omega\!\leq\!1J$).  The matrix elements are averaged over a small frequency window (of width $0.05J/\hbar$) for visualization.
	We used $\phi\!=\!2/7$ and $\varepsilon_F\!=\!-1.5J$.}
\label{smfig:CI_singleChiral}
\end{figure} 

We here study a fermionic, non-interacting CI state at $\phi\!=\!2/7$-flux per plaquette and consider the system being confined by a circular box of radius $r_{\rm wall}$, by using a hard-wall potential $V_{m,n}$. 
We fill all the single particle states up to the Fermi energy $\varepsilon_F\!=\!-1.5J$, lying in the first bulk gap. 
We show in Fig.~\ref{smfig:CI_singleChiral}(a) the matrix elements of a probe carrying angular momentum $\ell$ (analogous to a Laguerre-Gaussian beam) as a function of the excitation energy $\hbar\omega$. These matrix elements  demonstrate the presence of a single gapless chiral branch of excitations, well separated from bulk modes. While a linearly dispersing branch of excitations can be seen at $\hbar\omega\rightarrow0$ for $\ell\geq 0$, no such response is observed when $\ell<0$, highlighting the single chirality present at low energies. On the other hand, at energies of the order of the bulk gap ($\approx 1.6 J$), transitions are well visible both at $\ell\geq 0$ and $\ell<0$.
The energy separation between bulk and edge modes is used to test the edge CD response: In Fig.~\ref{smfig:CI_singleChiral}(b) the features sketched in Fig.~{\color{red!95!black}1} and described in the main text can be recognized and we can indeed extract the Chern number from the (numerically calculated) low-energy matrix elements $M_{\pm}^{n,0}$.
At low energy, in line with the circular symmetry of the boundary  and the presence of a single chiral edge mode, only the matrix elements $M_{+}^{n,0}$ are non-vanishing, as we illustrate in the inset of Fig.~\ref{smfig:CI_singleChiral}(b).
To extract the expected $|C|\approx1$, the area $A$ of the system needs to be carefully factored in, which is given by the total area $A\!=\!\pi r^2_{\rm wall}$.

\begin{figure}
\includegraphics[width=1\linewidth]{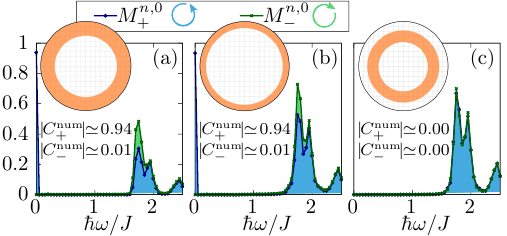}
\caption{Matrix elements $M_{\pm}^{n,0}$ for a CD probe that acts locally on an annular region of width $w$ (insets), for the same system parameters [Fig.~\ref{smfig:CI_singleChiral}]. The local probe with (a) $w\!=\!8$ and (b) $w\!=\!4$ couples to the chiral edge mode at low energies. The Chern number  
	$C_{\text{MB}}^{\rm num}$ extracted from the low-energy response ($\hbar\omega\leq J$)  remains quantized. 
	(c) A probe localized deep in the bulk yields a vanishing low-energy CD response.}
\label{smfig:mask}
\end{figure} 

As we discuss in the main text, the edge CD arises from a tiny region at the system's edge, which opens the possibility of measuring the Chern number from locally addressing the boundary of the system. We test this idea of local CD spectroscopy in Fig.~\ref{smfig:mask}, where we employ a spatial mask with an annular shape described by $f(r)=r\,\Theta_H(r-R_1)\Theta_H(R_2-r)$ to isolate the bulk and edge responses, where, $\Theta_H$ is the Heaviside-step function and we varied $R_1$, $R_2$ such that either both bulk and edge are addressed, or only one of them.
In particular, in Fig.~\ref{smfig:mask}(a) we illuminate both the edge and a substantial part of the bulk by choosing $R_1=r_{\rm wall}-8$ and $R_2=r_{\rm wall}$;
in Fig.~\ref{smfig:mask}(b) we probe the edge and a less extended portion of the bulk, by choosing $R_1=r_{\rm wall}-4$ and $R_2=r_{\rm wall}$. It can be seen that the edge CD response is not quantitatively modified, proving that the response is indeed strongly localized at the edge.
In Fig.~\ref{smfig:mask}(c) the probe is instead localized in the bulk, $R_1=r_{\rm wall}-13$ and $R_2=r_{\rm wall}-5$, and the edge is excluded.
It can be seen that in this case the edge CD drops to zero at low energies, since the edge of the system has not been targeted.
We emphasize that in all cases considered in Fig.~\ref{smfig:mask}, the numerically extracted Chern number is calculated by using the full system area $A\!=\!\pi r^2_{\rm wall}$, as $f(R)\sim R$.

\begin{figure}[t]
\includegraphics[width=1\linewidth]{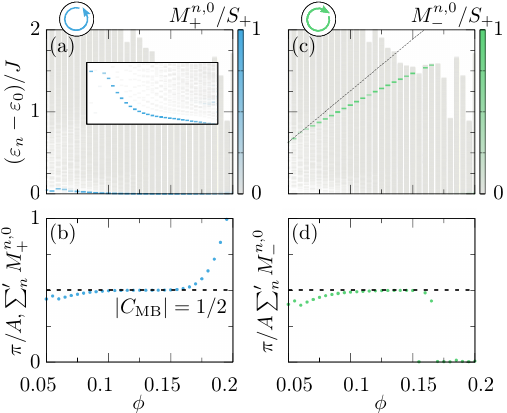}
\caption{
	(a) Energy spectrum for $N\!=\!3$ hard-core bosons on a $12\times12$ square lattice, in the presence of a weak harmonic confinement of strength $U_h\!=\!2\times10^{-3} J$, as a function of the flux $\phi$.
	The points have been colored according to the magnitude of the matrix elements $M^{n,0}_+$. 
	The inset shows a close-up on the low-energy features.
	(b) The matrix elements have been summed up to a cutoff energy $\sim 0.5 J$ to extract the $S_+$ transition rate; the primed summation is here a short-hand for this energy cutoff.
	The exact-diagonalization results are compared to the theory prediction (black dashed line).
	(c,d) Same but for the matrix elements $M^{n,0}_-$, associated with the bulk response. In (d), the primed-summation involves all the states above the cutoff $\sim0.5 J$.
	Black-dashes in (c) show the continuum-limit cyclotron gap $\hbar\omega_c/J\!=\!4\pi\phi$.
}
\label{smfig:FCI_N=3}
\end{figure} 

\subsubsection{Harper-Hofstadter FCI}
The addition of interactions to the Harper-Hofstadter Hamiltonian Eq.~\eqref{eq:Hofst.Hamiltonian} is known to allow for a plethora of FCI states, depending on the nature of the interactions and particle statistics~\cite{Sorensen,Hafezi_FQH,Palmer_FQH,Moore_edge_FCI,moller2009composite,Scaffidi_Hof,Roy_Hofstadter,Moller_higher,Murthy_Hof,Raciunas_18_PRA,Repellin_PRA,Palm_PRB,Bosl}.
We focus on bosons with a strong two-body Hubbard repulsion ($H_{\rm int}=\frac{U}{2}\sum_{m,n} \hat n_{m,n}(\hat n_{m,n}-1)$), which favors Laughlin-type FCI states. 
We confine the system by a harmonic potential $V_{m,n}=U_h(m^2+n^2)$, which helps stabilizing a Laughlin-like state over a broad interval of the flux per plaquette $\phi$, and makes the edge approximately circular in shape.

In the main text, numerical data for a system with $N=2$ particles on a $12\times12$ square lattice  have been presented as a function of $\phi$; 
extracting a large-enough portion of the spectrum (including the states which are bright to the counter-rotating probe $\propto x-iy$) for a larger system quickly becomes numerically challenging. 
We present in Fig.~\ref{smfig:FCI_N=3} data for $N=3$ particles on the same lattice. 
In panels (a,c) the matrix elements $M_{\pm}^{n,0}$ are displayed with color on top of the system's eigenenergies, while 
in panel (b)(/(d)) we show how the relevant edge(/bulk) transition rates can be extracted by summing these matrix-elements below(/above) a cutoff frequency ($\sim 0.5J/\hbar$). 
The same qualitative features discussed in the main text, as well as the same features shown in Fig.~{\color{red!95!black}2} for a smaller system, can be recognized. This supports the idea that these are not finite-size effects but indeed genuine physical effects.
However, the high-energy part at large fluxes $\phi$ suffers from numerical limitations: it is indeed ``white" because a limited number of low-energy eigenstates ($\approx 2000$) has been numerically obtained at every $\phi$. In the high-field region at high-energy, finite size effects are indeed visible (cf. Fig.~\ref{smfig:FCI_N=3}(c-d)) as evidence by the disappearence of the relevant transition line $M_-^{n,0}$ (green colored points).

\begin{figure}[t]
\includegraphics[width=1\linewidth]{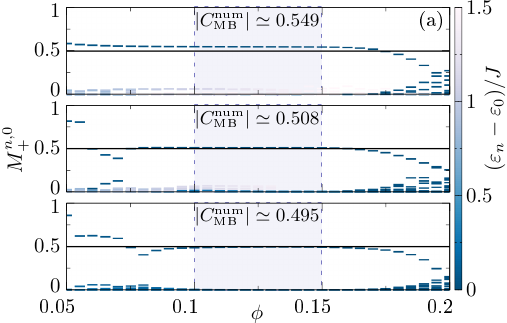}
\caption{
	Matrix elements ${M}^{n,0}_+$ for $f(r)\!=\!1$ (see text) as a function of the flux $\phi$; points are colored according to the energy of the $n$-th state, relative to the ground state. We consider
	(a) $N\!=\!2$, (b) $N\!=\!3$ and (c) $N\!=\!4$ bosons on a $12\times12$ lattice, in the presence of a weak harmonic confinement of strength $U_h\!=\!0.002 J$. 
	The points are compared with the theoretical prediction $\Delta S\!=\!|C_{\rm MB}|$ (black line).
	The reported values $|C_{\rm MB}^{\rm num}|$ are averages in the blue-shaded region.
}
\label{smfig:FCI_uniform}
\end{figure} 

We then test the idea of the local probe in this FCI setup. Since the system is small and it is challenging to spatially separate bulk and edge, we choose a generalized CD probe with a constant radial profile $f(r)=1$.
Such a probe has the advantage of providing direct access to the many-body Chern number, without any prior knowledge of the system's size.
We also note here that this probe is unphysical at $r\!=\!0$, but this does not affect the results since the bulk remains inert at low-energy: 
if the system is large enough, modifying the probe profile at $r\approx0$ (for example choosing $f(r)=\Theta_H(r-R)$) will yield the same results.
We display the relevant matrix elements $M_{\pm}$ in Fig.~\ref{smfig:FCI_uniform}, as a function of the flux, for various system sizes. Deep in the FCI region, a single low-energy edge mode is shown to react to the generalized CD probe, and leads to the quantized response $\Delta{S}\propto-C_{\rm MB}$. As we increase the number of particles, the value $|C_{\rm MB}^{\rm num}|$ that we extract from the plateau region saturates towards $|C_{\rm MB}|=1/2$. These results indicate how the many-body Chern number of correlated insulators can be directly extracted from the edge response using a Laguerre-Gaussian-type probe~\cite{Goldman_LG_2012,Binanti_edge}.

\begin{figure}
\includegraphics[width=1\linewidth]{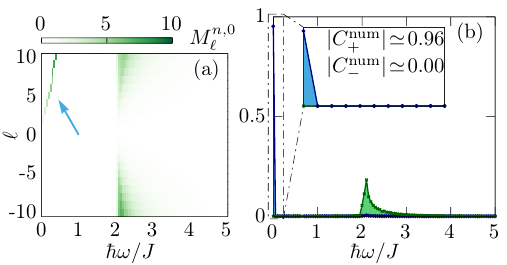}
\caption{
	(a) Angular-momentum-resolved spectroscopy in the single-particle Haldane model Eq.~\eqref{eq:haldane_hamiltonian}:~Matrix elements $M_{\ell}^{n,0}\!=\!\left|\int d^2\mathbf{r}\,  e^{i \ell \theta}\braket{n|\hat{\rho}(\mathbf{r})|0}\right|^2$, as a function of frequency, $\omega_{n,0}\!=\!(\varepsilon_n-\varepsilon_0)/\hbar$, and transferred (optical) angular momentum $\ell$; $\varepsilon_n$ is the energy of the $n$-th eigenstate.
	The presence of a single edge mode is highlighted by the arrows.
	(b) CD matrix elements $M_{\pm}^{n,0}$ [Eq.~{\color{red!95!black}5}], as a function of $\omega_{n,0}$, with the inset zooming in on the low-energy features. The numerically extracted Chern number contributions, $C_{\text{MB}}^{\rm num} = C_+^{\rm num}-C_-^{\rm num}$, are calculated by isolating the low-energy response ($\hbar\omega\!\leq\!1J$).  The matrix elements are averaged over a small frequency window (of width $0.05J/\hbar$) for visualization.
	We used $\phi_*\!=\!\pi/2$, $J'/J=0.2$ and $\varepsilon_F\!=\!0$. A circular hard-wall potential at $r_{\rm wall}=29.5a$ ($a$ being the nearest-neighbor distance) has been used.}
\label{smfig:haldane}
\end{figure} 

\subsubsection{Haldane CI}
We test our ideas in the case of the prototypical example of a QH insulator in the absence of a magnetic field: the Haldane model~\cite{Haldane88_PRL} on the honeycomb lattice, described by
\begin{equation} \label{eq:haldane_hamiltonian}
H=-J\sum_{\langle i,j\rangle} c_i^\dagger  c_j^\nodagger - J'\sum_{\langle\langle i,j\rangle\rangle} e^{\pm i \phi_*}c_i^\dagger  c_j^\nodagger + \sum_i V_i c^\dagger_i c_i^\nodagger
\end{equation}
where $\hat c_i^{(\dagger)}$ are annihilation (creation) operators at site $i$, 
$\langle i,j\rangle$ denote summation over the nearest neighbor sites, 
$\langle\langle i,j\rangle\rangle$ over the next to nearest ones,
and the phase $e^{\pm i \phi_*}$ has the positive (/negative) sign for hopping in the anticlockwise (/clockwise) direction.
Finally, the confinement term $V_i$ encodes a circular confining box-potential with hard walls at radius $r_{\rm wall}$, which defines a system with area $A\!=\!\pi r^2_{\rm wall}$.
The ground state of the Haldane model at half filling features two topologically non-trivial regions with Chern number $C=\pm1$ in addition to a trivial phase, where the value of the Chern number can be tuned by changing the tunneling strengths $J'/J$, the complex phase $\phi_*$ (encoding a staggered magnetic field) and sublattice energy offsets (here, we assumed the latter to vanish). 
In Fig.~\ref{smfig:haldane}(a) we plot the matrix elements of a probe carrying angular momentum $\ell$ (analogous to a Laguerre-Gaussian beam) as a function of the excitation energy $\hbar\omega$, demonstrating the presence of a single gapless chiral branch of excitations, well separated from bulk modes.
Fig.~\ref{smfig:haldane}(b) shows how the edge CD response successfully reveals the bulk Chern number. Upon isolating the low-frequency contributions, we obtain the numerically extracted Chern number $|C^{\rm num}|\approx0.96$.

\subsection{Multi-edge, circular boundary configurations}
We now present a broader discussion on the numerical results presented in the main text concerning multi-edge scenarios.
We first discuss more extensively the multi-edge CI results based on the Harper-Hofstadter model introduced before;
we then briefly introduce the Halperin $(2,2,1)$ state and discuss more extensively our numerical results.

\subsubsection{Non-interacting CI}
We consider here again the Harper-Hofstadter model Eq.~\eqref{eq:Hofst.Hamiltonian};
we study a fermionic, non-interacting CI state at $\phi\!=\!2/7$-flux per plaquette.
We fill all the single particle states up to the Fermi energy $\varepsilon_F\!=\!-1.5J$, lying in the first bulk gap. 
As we mentioned in the main text, instead of considering just a circular box of radius $r_{\rm wall}$ confining the system, we include in Fig.~{\color{red!95!black}3} a modified confining potential $V_{m,n}$ which, in addition to the hard-wall part at $r_{\rm wall}$, features a Gaussian bump of the form $V_{0}\,e^{-(\br-\br_0)^2/2\sigma_0^2}$ with width $\sigma_0=0.5$, centered close to the hard-wall edge at $r_0=r_{\rm{wall}}-2$. 
This Gaussian potential, if large enough, introduces an extra pair of edge modes of opposite chirality, localized on the ``sides" of the Gaussian bump -- the classical analog being skipping orbits following equipotential lines with velocity $\partial_r V(r)$, which takes opposite signs on opposite sides of the bump.
The idea, sketched in Fig.~{\color{red!95!black}3}(a), is corroborated by analyzing the matrix elements of a probe carrying angular momentum $\ell$ in Fig.~{\color{red!95!black}3}(b), analogously to what is shown in Fig.~\ref{smfig:CI_singleChiral}. In this case the presence of a counter-propagating mode at low-energy can clearly be seen by the bright response at $\hbar\omega\approx 0$ when $\ell<0$.
Since this bump alters the system area (the ground state density being depleted at the bump region), when extracting the Chern number we heuristically compensate for the depletion by considering an effective system's area $A\!=\!\pi r^2_{\rm wall}-A_{\rm bump}$; namely, we subtract the bump's area, computed by integrating the gaussian shape of the bump $A_{\rm bump}=\int d^2\mathbf{r}\,e^{-(\br-\br_0)^2/2\sigma_0^2}$, from the area enclosed by the hard wall.

\subsubsection{Halperin $(2,2,1)$}
\begin{figure}
\includegraphics[width=1\linewidth]{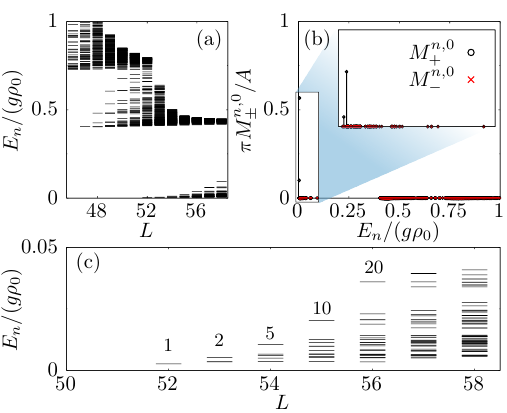}
\caption{
	(a) Spectrum of a rapidly-rotating, circularly symmetric cloud of two-component bosons, interacting via Eq.~\eqref{eq:halperin221interaction}.
	At $L=L_{(2,2,1)}=52$, the incompressible Halperin $(2,2,1)$ state appears, together with a branch of chiral low-energy edge modes at $L>L_{(2,2,1)}$. 
	Several other states can be seen at any $L$, separated from the ground-state by an energy scale of the order of the interaction energy $g \rho_0$ (see Eq.~\eqref{eq:halperin221interaction}).
	(b) CD matrix elements $M_{\pm}^{n,0}$, as a function of the eigenstate energy.
	Only the $M_{+}^{n,0}$ can be seen to be non-zero within the energy window we accessed numerically, and, crucially, only for the low-lying edge modes.    
	The inset displays a close-up view on this low-energy part.
	(c) Zoom on the low-energy part of the spectrum displayed in panel (a). The numbers (displayed up to $L-L_{(2,2,1)}=4$) denote the counting of the quasi-degenerate edge states~\cite{myNote}. 
}
\label{smfig:sm_halperin_full}
\end{figure} 

As a paradigmatic example relevant for atomic gases in rapid rotation, we consider the incompressible bosonic Halperin $(2,2,1)$ state, which appears as the densest ground state for a strongly-interacting two-component system under strong perpendicular magnetic fields.
This ground-state is found at large values of the total angular momentum~\cite{Cooper_2008} $L_{\rm (2,2,1)}=N_A(N_A-1)+N_B(N_B-1)+N_A N_B$; it is not difficult to show that the bulk filling fraction of such a state is $\nu=|C_{\rm MB}|=2/3$.
Such an incompressible state hosts at his boundary two copropagating edge modes~\cite{Wen_AdvPhys_1995}, which we are going to numerically analyze in the following.

More explicitly, we consider rapidly-rotating two-component bosons
\begin{equation}
\label{eq:kineticHamiltonian}
H_0 = \sum_{\sigma=A,B}\sum_{i=1}^{N_\sigma} \frac{\boldsymbol{\pi}^2_{i,\sigma}}{2m}
\end{equation}
where $N_A$(/$N_B$) is the number of $A$(/$B$) bosons and $\boldsymbol{\pi}_{i,\sigma}=\mathbf{p}_{i,\sigma}-q\mathbf{A}(\mathbf{r}_{i,\sigma})$ the kinetic momentum operator due to the (synthetic) gauge field $q\mathbf{A}$.
This Hamiltonian is well-known to be a harmonic oscillator one, with an energy spectrum given by Landau levels with energies $\hbar \omega_C\left(n_{\rm LL}+\frac{1}{2}\right)$, $\omega_C=qB/m$ being the cyclotron frequency and $n_{\rm LL}\in\mathbb{N}$ the Landau level index.

We consider the particles to be interacting via purely contact interactions, 
\begin{equation}
\label{eq:halperin221interaction}
\begin{split}
	V_{\rm (2,2,1)} = 
	\sum_{\sigma=A,B}&g_{\sigma,\sigma}\sum_{i<j}\delta^{(2)}({\bf r}_{\sigma,i}-{\bf r}_{\sigma,j}) + \\ +&g_{A,B}\sum_{i,j}\delta^{(2)}({\bf r}_{A,i}-{\bf r}_{B,j}).
\end{split}
\end{equation}
We will restrict ourselves to the case $g_{\sigma,\sigma'}=g>0$ for any possible values of $\sigma,\sigma'$, and project the dynamics to the lowest Landau level of Eq.~\eqref{eq:kineticHamiltonian}, given by $n_{\rm LL}=0$.
In order to make the edge modes dispersive we also include a strongly-anharmonic circularly symmetric ring-shaped box potential $V_{\rm w}(r) = u_{\rm w} \Theta_H(R_{\rm w}-r)$,~\cite{Macaluso_PRA_2017}
\begin{equation}
H_{(2,2,1)} =  P_{\rm LLL} \left(V_{\rm (2,2,1)}+ \sum_{\sigma}\sum_{i=1}^{N_\sigma}V_{\rm w}(r_i)\right)  P_{\rm LLL},
\end{equation}
where $P_{\rm LLL}$ is the projector onto the lowest Landau level and the constant energy offset $(N_A+N_B)\hbar\omega_C/2$ associated to the lowest-Landau level zero-point motion has been disregarded.
We diagonalize this Hamiltonian numerically, labeling the energy modes with the angular momentum eigenvalue.
We report here numerical results for the largest system we diagonalized using  exact diagonalization techniques, $N_A=5$, $N_B=4$.
We considered $u_{\rm w}=0.1 g / (\hbar\omega_C l_B^2)$ and $R_{\rm w}/l_B=2\sqrt{N_A+N_B}+1$, where $l_B=\sqrt{\hbar/qB}$ is the magnetic length.

In Fig.~{\color{red!95!black}4}(a) we show a radial cut of the system's density. It can be seen that the characteristic featureless plateau at $\rho_{0}=\frac{|C_{\rm MB}|}{2\pi l_B^2}$ starts forming in the bulk already at this system size.
In Fig.~\ref{smfig:sm_halperin_full}(a) we display the  spectrum as a function of the conserved angular momentum, where an almost zero-energy non-degenerate ground state is visible at $L=L_{(2,2,1)}$. 
It can be seen that low-energy excitations are available only for $L\geq L_{(2,2,1)}$, explicitly showing the chirality of the edge modes:
for $L<L_{(2,2,1)}$, no state is present at low-energy. 
When $L>L_{(2,2,1)}$ several low-energy states do appear. Their counting ($1,2,5,10,20,\dots$)  -- see Fig.~\ref{smfig:sm_halperin_full}(c) -- matches that expected from the presence of two branches of co-propagating bosonic chiral modes~\cite{myNote}.
Several other states can be seen at every $L$ at energies of the order of the typical interaction energy $g\rho_0$, corresponding to gapped excitations of the Halperin $(2,2,1)$ state.

Since we are dealing with a circularly symmetric state hosting two co-propagating edge modes, we expect the edge CD to be associated purely to the $x+iy$ probe, while the response to the $x-i y$ one should be suppressed at low energy. 
In other words, we only expect the matrix elements $M_+^{n,0}$ to be non-vanishing at low energies, with no contribution coming from the $M_{-}^{n,0}$ terms.
This picture is confirmed in Fig.~{\color{red!95!black}4}(b), where we show that the matrix elements $M_{\pm}^{n,0}$ associated to the low-energy edge modes completely saturate the sum rule Eq.~\eqref{eq:edge_dichro}.
More explicitly, in Fig.~\ref{smfig:sm_halperin_full}(b) we show how, within the portion of the energy spectrum we were able to access via exact diagonalization techniques, $M_{-}^{n,0}$ is negligible while $M_{+}^{n,0}$ is non-zero only below the many-body energy gap $\propto g\rho_0$, i.e.~in correspondence of the chiral gapless edge modes.

\subsection{Single-edge, deformed boundaries}
Finally, we present numerical results concerning the deformed-boundary scenario. 
Beside providing further details on how the numerical calculations presented in the main text were performed, 
we present some additional numerical results which strengthen our arguments but are not necessary for understanding the material contained in the main document.
In particular, we begin with an elliptically deformed CI in the Harper-Hofstadter model; we then discuss the case of an elliptically confined Laughlin FQH state, and we conclude with a more complicated ``clover"-like deformation of a non-interacting IQH state at unit filling.

\begin{figure}[t]
\includegraphics[width=1\linewidth]{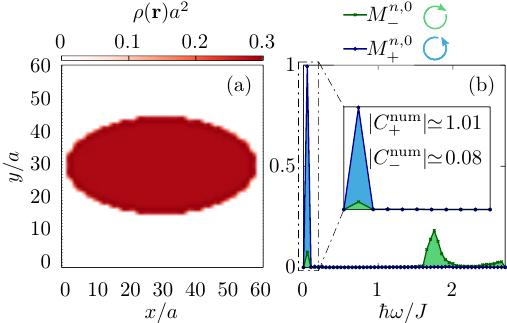}
\caption{(a) Ground state density for a non-interacting fermionic CI in the Harper-Hofstadter model, confined by an elliptically-shaped box; the ratio between the two semiaxis of the trap, $\Delta x$ and $\Delta y$, was fixed to $\Delta x/\Delta y=1.97$. 
	The flux per plaquette has been fixed to $\phi=2/7$, and the Fermi energy approximately in the middle of the first gap, $\varepsilon_F\!=\!-1.5J$. 
	(b) CD matrix elements $M_{\pm}^{n,0}$ [Eq.~{\color{red!95!black}5}], as a function of the excitation frequency $\omega_{n,0}$, with the inset zooming in on the low-energy features. The numerically extracted Chern number contributions, $C_{\text{MB}}^{\rm num} = C_+^{\rm num}-C_-^{\rm num}$, are calculated by isolating the low-energy response ($\hbar\omega\!\leq\!1J$).}
\label{fig:sm_CI_elliptical}
\end{figure}

\subsubsection{Elliptically deformed non-interacting CI}

We show in Fig.~\ref{fig:sm_CI_elliptical} the results obtained in the case of a Harper-Hofstadter non-interacting fermionic CI on the square lattice, obtained by numerically diagonalizing the Harper-Hofstadter Hamiltonian Eq.~\eqref{eq:Hofst.Hamiltonian} at flux $\phi\!=\!2/7$,
in the presence of an elliptical box potential
$V_{m,n} = 0$ if $m^2/a^2+n^2/b^2 \leq 1$, and infinity otherwise. We then filled the single-particle energy states up to $\epsilon_F=-1.5J$, lying in the first bulk gap.
We chose $a=29.5$ and $b=15$ in units of the lattice constant.

It can be seen that the numerical results we present here do agree with the expected qualitative behaviour described in the main text. 
In particular, (i) $S_-$ does not vanish; (ii) the difference $S_+-S_-$ is quantized in terms of the many-body Chern number, in agreement with Eq.~\eqref{eq:dichroism_arbitrary_shape}; (iii) the ratio $S_-/S_+\approx0.08$ is accurately described by Eq.~{\color{red!95!black}6}; $\lambda$ can indeed be obtained from the ratio between the two semi-axes of the ellipse according to Eq.~\eqref{eq:boundary_shape_ellipse}. For the elliptical box confinement we considered, $\frac{\Delta x}{\Delta y}=e^{-2\lambda}\simeq 1.97$; Eq.~{\color{red!95!black}6} then gives $S_-/S_+\simeq 0.1$, which approximately agrees with the value we numerically extract.
We attribute the difference to finite size effects on the lattice, but postpone more extensive and systematic investigation to future studies.

\begin{figure}[t]
\includegraphics[width=1\linewidth]{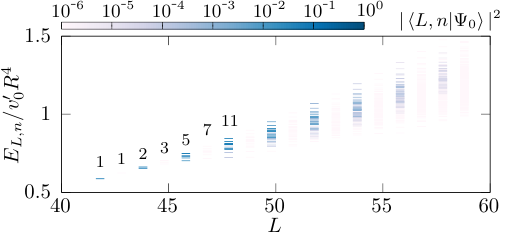}
\caption{Overlaps between the ``elliptic" Laughlin state $\ket{\Psi_0}$ (i.e. the ground state in the presence of the anisotropic and anharmonic confinement) and the eigenstates $\ket{L,n}$ of Eq.~\eqref{eq:sm_laughlin_pseudop} in the presence of an additional (very weak) isotropic anharmonic trap $v_0' r^4$, introduced to lift the edge-modes degeneracies;  the numbers (displayed up to $L-L_0=6$) denote the counting of the quasi-degenerate states, expected from the presence of a single chiral mode~\cite{myNote}.}
\label{fig:sm_eigenvectors_weights}
\end{figure} 

\subsubsection{Elliptically deformed Laughlin state}
We also considered the case of a bosonic Laughlin fractional quantum Hall state at $\nu=1/2$ filling in a synthetic magnetic field, stabilized by strong contact interactions~\cite{Cooper_2008} $V_{1/2}=g \sum_{i<j}\delta^{(2)}({\bf r}_i-{\bf r}_j)$, relevant for rapidly rotating atomic gases~\cite{Fletcher_Science_2021}.
Analogously to the Halperin $(2,2,1)$ case discussed above, we study the strong $B$-field limit, so that the dynamics can be restricted to the lowest Landau level $n_{\rm LL}=0$,
\begin{equation}
\label{eq:sm_laughlin_pseudop}
H_{1/2} =  P_{\rm LLL} V_{1/2} P_{\rm LLL}.
\end{equation}
Moreover, to showcase the generality of our results, instead of confining the system with the elliptical harmonic trap Eq.~\eqref{eq:elliptic_potential}, we apply an anharmonic one $U_{2,\lambda}=v_0 (e^{2\lambda}x^2+e^{-2\lambda}y^2)^2$. 
We consider it to be weak enough so that coupling to the states above the many-body gap $\Delta\approx g n_{2D} = \frac{g}{4\pi l_B^2}$ can be safely neglected: if the confinement at the edge position $r\appropto\sqrt{N}$ becomes too strong, it will be energetically convenient to condense quasiparticles in the bulk, paying an interaction cost $\Delta$, so as to shrink the cloud.
In line with these ideas, we first diagonalize the rotationally-invariant lowest Landau level projected interaction $H_{1/2}$, labelling the eigenenergies by the conserved angular momentum quantum number $L$~\cite{Cooper_2008}. 
We then restrict to the subspace of states laying below the many-body gap $\Delta$ so as to diagonalize the anisotropic confinement part $U_{2,\lambda}(x,y)$.
This procedure is numerically convenient because it significantly reduces the Hilbert space dimension, making the numerical exact diagonalization feasible.

The spectrum of a $N=7$ system in the {\it absence} of the anisotropic confinement is shown in Fig.~\ref{fig:sm_eigenvectors_weights}. For graphical purposes, in order to lift the degeneracy of the low-lying states~\cite{Wen_AdvPhys_1995, CooperSimon_PRL_2015}, an extremely weak and isotropic confinement $U_{2,\lambda=0} = v_0' r^4$ ($v_0'\ll v_0$) has also been introduced. The points are colored according to the overlap between the corresponding eigenstate and the ground state of the system in the {\it presence} of the anisotropic confinement $U_{2,\lambda}(x,y)$. It can be seen that the new ground state retains a significant overlap with the unperturbed Laughlin state at $L_0=N(N-1)=42$, but gets contributions from states with $L-L_0=0\pmod 2$ (due to the $180^\circ$ rotation symmetry of the trap).
The density of the anisotropic ground state for a selected value of $\lambda=0.2$ is shown in Fig.~{\color{red!95!black}5}(a).  
In Fig.~{\color{red!95!black}5}(b) we show that $\Delta S$ is robustly quantized in units of the many-body Chern number $|C_{\rm MB}|=1/2$, corresponding to a bosonic Laughlin state, as the ellipticity parameter $\lambda$ is varied. On the other hand $S_-/S_+$ as a function of $\lambda$, plotted in Fig.~{\color{red!95!black}5}(c), reflects the cloud's anisotropy according to Eq.~{\color{red!95!black}6} with a very good degree of accuracy.

\begin{figure}[t]
\includegraphics[width=1\linewidth]{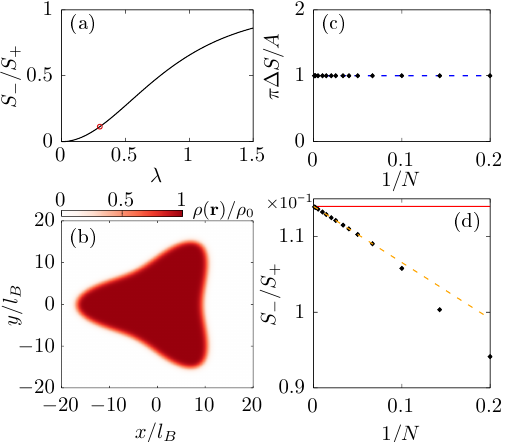}
\caption{(a) Numerically computed \cll value of the ratio of $S_-/S_+$ as a function of the trap deformation factor $\lambda$, for $k=3$ (see Eq.~\eqref{eq:confinement_deformed}). 
	The red circle identifies the value $\lambda=0.3$ used in the other panels.
	(b) Ground state density $\rho(\mathbf{r})$ for an integer quantum Hall system of $N=80$ fermions in the non-circularly symmetric potential in Eq.~\eqref{eq:confinement_deformed}. The density $\rho(\mathbf{r})$ has been normalized to the bulk one, $\rho_0=\frac{1}{2\pi l_B^2}$.
	(c) The circular dichroic response $\Delta S$ as a function of the number $N$ of particles in the two-dimensional trap.
	(d) The ratio $S_-/S_+$ as a function of the number of particles is compared with the value computed in panel (a). The red line indicates the theoretical prediction $S_-/S_+\simeq0.114$. 
	The numerical values (black points) approach it $\propto 1/N$ as the size of the system is made larger; in the scaling limit (orange dashed line) we extract (from a linear fit) $S_-/S_+\simeq0.114$, in accordance with the theory.}
\label{fig:sm_IQH_k=3}
\end{figure} 

\subsubsection{``Clover"-shaped non-interacting integer quantum Hall system}

We finally consider a potential of the form Eq.~\eqref{eq:confinement_deformed}, with $k=3$, whose equipotential lines draw a ``clover" shape (with three-fold rotation symmetry).
In this case, to the best of our knowledge, it is not possible to rewrite Eq.~\eqref{eq:x_y_at_the_edge} in simple terms. 
However, as anticipated, these relations can be used to numerically compute the ratio $S_-/S_+$ (through the use of Eq.~\eqref{eq:Spm_deformed}) 
and compare it with numerical results obtained by performing exact-diagonalization calculations on a finite system. 
For simplicity, we restrict our attention to a system of non-interacting fermions in the lowest-Landau level $n_{\rm LL}=0$, confined by Eq.~\eqref{eq:confinement_deformed} with $k=3$.

In Fig.~\ref{fig:sm_IQH_k=3}(a), we show the numerically computed theoretical ratio $S_-/S_+$ as a function of $\lambda$. Notice again how the theoretical ratio depends purely on the edge geometry and not on the system size.
In Fig.~\ref{fig:sm_IQH_k=3}(b), we show the ground state density $\rho(x,y)$ for $N=80$ non-interacting fermions, for a fixed value of $\lambda=0.3$. 
For the same value of $\lambda$, in Fig.~\ref{fig:sm_IQH_k=3}(c,d) we show  that $\Delta S$ is quantized as a function of the number of particles $N$ in accordance with Eq.~\eqref{eq:dichroism_arbitrary_shape} and that the theoretical prediction for the ratio $S_-/S_+$ is approached as the system is made larger.

\end{supplementalMaterials}


\begin{thebibliography}{95}
\bibitem [{myN()}]{myNote}%
	\BibitemOpen
	\href@noop {} {}\bibinfo {note} {The counting of low-energy edge excitations
		can be obtained as follows. In the case of a single chiral edge mode
		(Fig.~\ref{smfig:sm_halperin_full}), all the states at angular momentum
		$\Delta L$ can be obtained by acting on the vacuum $\ket{0}$ with the
		operators $\hat b_\ell^\dagger$ of
		Eq.~\eqref{eq:single_edge_bosonization}:~the total number of such states is
		given by the number of partitions of $\Delta L$, namely the number of ways of
		dividing up the angular momentum $\Delta L$ into independent sets
		$[\ell_1,\ell_2,...]$ that satisfy $\sum_i \ell_i=\Delta L$; we denote such a
		number as $p(\Delta L)$. In the case of two co-propagating chiral modes
		(Fig.~\ref{fig:sm_eigenvectors_weights}), the angular momentum $\Delta L$ can
		split in between the two branches. Therefore, the total number of excitations
		at angular momentum $\Delta L$ is $\sum_{\delta l=0}^{\Delta L} p(\Delta
		L-\delta l) p(\delta l)$. In general, finite size corrections in the counting
		eventually emerge at large $\Delta L$ due to the finite number of particles
		making up the fractional quantum Hall liquid.}\BibitemShut {Stop}%
\end{thebibliography}
\end{document}